\documentclass[leqno, 10pt]{article}
\usepackage{times}
\usepackage{amsmath}
\usepackage{amssymb}
\usepackage{setspace}

\numberwithin{equation}{section}

\newtheorem{theorem}{Theorem}[section]
\newtheorem{definition}[theorem]{Definition}
\newtheorem{lemma}[theorem]{Lemma}
\newtheorem{corollary}[theorem]{Corollary}

\begin{document}

\date{}

\title{Denotational semantics for modal systems S3--S5 extended by axioms for propositional quantifiers and identity}

\author{Steffen Lewitzka\thanks{Universidade Federal da Bahia -- UFBA,
Instituto de Matem\'atica,
Departamento de Ci\^encia da Computa\c c\~ao,
Campus de Ondina,
40170-110 Salvador -- BA,
Brazil,
e-mail: steffen@dcc.ufba.br}}

\maketitle

\begin{abstract}
There are logics where necessity is defined by means of a given identity connective: $\square\varphi := \varphi\equiv\top$ ($\top$ is a tautology). On the other hand, in many standard modal logics the concept of propositional identity (PI) $\varphi\equiv\psi$ can be defined by strict equivalence (SE) $\square(\varphi\leftrightarrow\psi)$. All these approaches to modality involve a principle that we call the Collapse Axiom (CA): ``There is only one necessary proposition." In this paper, we consider a notion of PI which relies on the identity axioms of Suszko's non-Fregean logic $\mathit{SCI}$. Then $S3$ proves to be the smallest Lewis modal system where PI can be defined as SE. We extend $S3$ to a non-Fregean logic with propositional quantifiers such that necessity and PI are integrated as non-interdefinable concepts. CA is not valid and PI refines SE. Models are expansions of $\mathit{SCI}$-models. We show that $\mathit{SCI}$-models are Boolean prealgebras, and vice-versa. This associates Non-Fregean Logic with research on Hyperintensional Semantics. PI equals SE iff models are Boolean algebras and CA holds. A representation result establishes a connection to Fine's approach to propositional quantifiers and shows that our theories are \textit{conservative} extensions of $S3$--$S5$, respectively. If we exclude the Barcan formula and a related axiom, then the resulting systems are still complete w.r.t. a simpler denotational semantics.
\end{abstract}


\section{Introduction}

The semantical approach to some Lewis-style modal logics studied in this paper relies on the principles of R. Suszko's non-Fregean logic (see, e.g., \cite{blosus1, blosus2, sus2, sus3}). The essential feature of a non-Fregean logic is an identity connective $\equiv$ such that $(\varphi\equiv\psi)\rightarrow(\varphi\leftrightarrow\psi)$ is a theorem but the so-called Fregean Axiom $(\varphi\leftrightarrow\psi)\rightarrow(\varphi\equiv\psi)$ is not valid. A formula $\varphi\equiv\psi$ can be read as ``$\varphi$ and $\psi$ have the same denotation." The basic non-Fregean logic is the Sentential Calculus with Identity $\mathit{SCI}$ \cite{blosus1, blosus2}. $\mathit{SCI}$ extends classical propositional logic by an identity connective and identity axioms which can be given by the following three schemes: $\varphi\equiv\varphi$, $(\varphi\equiv\psi)\rightarrow(\varphi\rightarrow\psi)$, and $(\varphi\equiv\psi)\rightarrow(\chi[x:=\varphi]\equiv\chi[x:=\psi])$, where $\chi[x:=\varphi]$ is the formula that results from substitutions of all occurrences of variable $x$ in $\chi$ with formula $\varphi$. A model of $\mathit{SCI}$ can be defined as a structure $\mathcal{M}=(M, \mathit{TRUE}, f_\bot, f_\top, f_\neg, f_\vee, f_\wedge, f_\rightarrow, f_\equiv)$ such that for all elements $a,b$ of the universe $M$, the conditions (ii)(a)--(e) and (g) of Definition \ref{200} below are satisfied. An assignment (or valuation) is a function $\gamma\colon V\rightarrow M$ from the set of propositional variables $V$ to $M$ which extends in the canonical way to a function from the whole set of formulas to $M$. The satisfaction relation then is defined as $(\mathcal{M},\gamma)\vDash\varphi$ $:\Leftrightarrow$ $\gamma(\varphi)\in TRUE$. More expressive non-Fregean logics which contain also propositional quantifies and further ingredients are studied, e.g., in \cite{sus2, blo, lewigpl}.

We define a proposition as the denotation $\gamma(\varphi)\in M$ of a formula $\varphi$ in a model $\mathcal{M}$ under a given assignment $\gamma$.\footnote{Suszko refers to the elements of a non-Fregean model as \textit{situations}. His aim was to develop a \textit{situational semantics} \cite{woj} as an attempt to formalize aspects of Wittgenstein's \textit{Tractatus} \cite{sus1}.} The proposition denoted by $\varphi$ can be identified with the equivalence class $\{\psi\mid (\mathcal{M},\gamma)\vDash\varphi\equiv\psi\}$ which is the set of all formulas having the same denotation as $\varphi$. We call this set the \textit{extension} of $\varphi$. An \textit{extensional model} contains only two propositions: the True and the False. In such a model, a proposition is given by its truth-value. If there were only extensional models, then the Fregean Axiom would be valid and $\mathit{SCI}$ would be equivalent with classical propositional logic. The \textit{intension} of a formula $\varphi$ is expressed by its syntactical form.\footnote{In a non-Fregean logic with propositional quantifiers, alpha-congruent formulas, i.e., formulas that differ at most on their bound variables, express the same intension and should denote the same proposition. We write $\varphi=_\alpha\psi$ if $\varphi$ and $\psi$ are alpha-congruent.} In a non-Fregean logic with propositional quantifiers we call a model \textit{intensional} if extension and intension of sentences (formulas with no free variables) can be put in one-to-one correspondence, i.e., if for all sentences $\varphi,\psi$, $(\mathcal{M},\gamma)\vDash\varphi\equiv\psi$ iff $\varphi=_\alpha\psi$. The existence of such a model (see \cite{lewigpl}) ensures that $\varphi\equiv\psi$ is logically valid iff $\varphi=_\alpha\psi$. That is, besides alpha-congruence, no further identifications between sentences are forced by the logic. Therein lies the expressive power of non-Fregean logic. Intensions of sentences are no longer indiscernible and semantic properties can be modeled easily (see, e.g., \cite{lewnd, lewsl, lewigpl}). This feature, however, can be lost if a specific non-Fregean theory involves too strong principles.

Early approaches to modality in logics with an identity connective have been developed by M. J. Cresswell \cite{cre1, cre2} and R. Suszko \cite{sus2}, see also the \textit{Historical Note} at the end of \cite{sus2}. Suszko elaborates two particular $\mathit{SCI}$-theories which correspond to the modal logics $S4$ and $S5$, respectively. Ishii \cite{ish1,ish2} is able to generalize these results by modifying the axioms of propositional identity of $\mathit{SCI}$. His system $\mathit{PCI}$ corresponds exactly to modal logic $K$. Moreover, he shows that $\mathit{PCI}$ can be extended to systems which correspond to many other normal modal logics, including $S4$ and $S5$. All these proposals have in common that the modal operator is introduced or defined by means of the identity connective: $\square\varphi :=\varphi\equiv\top$. Consequently, there is only one necessary proposition, namely the proposition denoted by $\top$. We call this principle the \textit{Collapse Axiom}. Moreover, propositional identity $\varphi\equiv\psi$ is given by strict equivalence $\square(\varphi\leftrightarrow\psi)$ and models are forced to be Boolean algebras (with some additional structure). In particular, logically equivalent formulas, such as  $\varphi\rightarrow\psi$ and $\neg\varphi\vee\psi$, are indiscernible although they express different intensions. We argue that these algebraic constraints are (at least in case of Lewis systems $S3$ -- $S5$ ) unnecessarily strong and restrict the potential of intensional modeling in non-Fregean logic. For instance, in \cite{lewsl} it is shown that if a non-Fregean model has many necessary (=known) propositions, then common knowledge in a group can be modeled in a natural way. The approaches mentioned above adopt the limitations which are already inherent in possible worlds semantics. In fact, if at a given normal world $w$ (in some Kripke frame), the proposition denoted by formula $\varphi$ is defined as the set of those worlds which are accessible from $w$ and where $\varphi$ is true, then $\varphi$ and $\psi$ denote the same proposition iff $\square(\varphi\leftrightarrow\psi)$ is true at $w$. Hence, propositional identity $\varphi\equiv\psi$ is given by strict equivalence. Suppose now $\square\varphi$ and $\square\psi$ are true at $w$. Since $\varphi\rightarrow(\psi\rightarrow\varphi)$ is a theorem, Necessitation yields $\square(\varphi\rightarrow(\psi\rightarrow\varphi))$. Applying the $K$-axiom and Modus Ponens, we derive $\square(\psi\rightarrow\varphi)$. Similarly, we obtain $\square(\varphi\rightarrow\psi)$. Thus, $\varphi$ and $\psi$ are strictly equivalent and denote the same proposition. Thus, the Collapse Axiom $(\square\varphi\wedge\square\psi)\rightarrow (\varphi\equiv\psi)$ is valid. One goal of this paper is to capture some Lewis modal systems by a non-Fregean semantics without the above described limitations. 
In particular, the Collapse Axiom should be invalid. Consequently, necessity and propositional identity must be axiomatized independently from each other. A further goal of this paper is to find an appropriate axiomatization of propositional quantifiers (i.e., quantifiers that range over the model-theoretic universe of a model) which is independent from specific properties of the possible worlds framework. In a first approach, we give an axiomatization which essentially corresponds to that presented by K. Fine \cite{fin} and which is sound and complete w.r.t. our first kind of denotational semantics. That axiomatization contains the Barcan formula, valid in the possible worlds semantics considered in \cite{fin, bul}, as well as a related extensional principle. Both principles can be excluded from the original axiomatization if we work with a weaker, simpler and in some sense ``more intensional" denotational semantics which we consider in the last section of the paper.

\section{The deductive system}

The set $Fm(C)$ of formulas is inductively defined over a set $V=\{x_0,x_1,x_2,...\}$ of propositional variables, a set $C$ of propositional constants such that $\top,\bot\in C$, logical connectives $\neg, \rightarrow, \vee, \wedge, \bot, \top$, the identity connective $\equiv$, the modal operator $\square$ for necessity and a universal propositional quantifier $\forall$. $\varphi\leftrightarrow\psi$ is an abbreviation for $(\varphi\rightarrow\psi)\wedge(\psi\rightarrow\varphi)$. By $var(\varphi)$, $fvar(\varphi)$, $con(\varphi)$ we denote the set of variables, free variables, constants occurring in formula $\varphi$, respectively. These notations also apply (in the obvious way) to sets of formulas $\Phi$, e.g., $fvar(\Phi)$ etc. A substitution is a function $\sigma\colon V\cup C\rightarrow Fm(C)$. If $u_1,...,u_n\in V\cup C$, $\psi_1,...,\psi_n\in Fm(C)$ and $\sigma$ is a substitution, then $\sigma[u_1:=\psi_1,...,u_n:=\psi_n]$ is the substitution which maps $u_i$ to $\psi_i$ ($i=1,...,n$) and coincides with $\sigma$ on $(V\cup C)\smallsetminus\{u_1,...,u_n\}$. The identity substitution $u\mapsto u$ is denoted by $\varepsilon$. Instead of $\varepsilon[u_1:=\psi_1,...,u_n:=\psi_n]$ we also write $[u_1:=\psi_1,...,u_n:=\psi_n]$. If we write $\sigma\colon V\rightarrow Fm(C)$, then we tacitly assume that $\sigma$ is a substitution satisfying $\sigma(c)=c$ for all $c\in C$. A substitution $\sigma$ extends to a function from $Fm(C)$ to $Fm(C)$ which we denote again by $\sigma$. We apply postfix notation: $\varphi[\sigma]$. The extension is defined canonically in most of the cases: $(\varphi\vee\psi)[\sigma]:=\varphi[\sigma]\vee\psi[\sigma]$, etc. Only the quantifier case needs a specification: 
\begin{equation*}
\begin{split}
(\forall x\varphi)[\sigma]=\forall y(\varphi[\sigma[x:=y]]),
\end{split}
\end{equation*}
where $y$ is the least variable of $V$ greater than all elements of $\bigcup\{fvar(\sigma(u))\mid u\in fvar(\forall x\varphi)\cup con(\forall x\varphi)\}$. We say that the variable $y$ is forced by the substitution $\sigma$ w.r.t. $\forall x\varphi$. 

In analogy to the Lambda Calculus, two formulas $\varphi,\psi$ are said to be alpha-congruent, notation: $\varphi=_\alpha\psi$, if $\varphi$ and $\psi$ differ at most on their bound variables. For instance, $\forall x((x\equiv\bot)\vee (x\equiv\top))=_\alpha\forall y((y\equiv\bot)\vee (y\equiv\top))$. Alpha-congruent formulas express the same intension and should denote the same proposition in every model. This is ensured by the model-theoretic semantics.

We assume that $\forall x\varphi\in Fm(C)$ implies $x\in fvar(\varphi)$. Strings such as $\forall x c$ or $\forall y(x\equiv x)$ are not formulas. This can be guaranteed by a suitable definition of $Fm(C)$, see \cite{lewigpl}. Also for a proof of the following fact we refer the reader to \cite{lewigpl}. Recall that $\varepsilon$ is the identity substitution. $\varepsilon$ applied to a formula may result in a renaming of bound variables.

\begin{lemma}[\cite{lewigpl}]\label{50}
Let $\varphi,\psi\in Fm(C)$. Then $\varphi[\varepsilon]=_\alpha\varphi$. Moreover, $\varphi=_\alpha\psi\Leftrightarrow\varphi[\varepsilon]=\psi[\varepsilon]$.
\end{lemma}

The quantifier rank $qr(\varphi)$ of a formula $\varphi$ is recursively defined in the following way: $qr(u)=0$ for $u\in V\cup C$, $qr(\neg\psi)=qr(\square\psi)=qr(\psi)$, $qr(\psi@\chi)=max\{qr(\psi),qr(\chi)\}$, where $@\in\{\vee,\wedge,\rightarrow,\equiv\}$, $qr(\forall x\psi)=1+qr(\psi)$.\\

A sentence is a formula with no free variables. $Fm_m\subseteq Fm(C)$ is the set of formulas of basic modal logic, i.e., the set of those formulas which are quantifier-free, do not contain the identity connective and do not contain constants distinct from $\bot,\top$. $Fm_p$ is the set of those formulas of $Fm_m$ which do not contain the modal operator $\square$, i.e., $Fm_p$ is the set of formulas of basic propositional logic. By a substitution-instance of $\varphi\in Fm_p$ we mean a formula which results from uniformly replacing some variables in $\varphi$ by formulas of $Fm(C)$.\\

All formulas of the following form are axioms:
\begin{enumerate}
\item propositional tautologies and their substitution-instances
\item $\square\varphi\rightarrow\varphi$
\item $\square(\varphi\rightarrow\psi)\rightarrow (\square\varphi\rightarrow\square\psi)$
\item $\square(\varphi\rightarrow\psi)\rightarrow\square (\square\varphi\rightarrow\square\psi)$
\item $\varphi\equiv\psi$, whenever $\varphi=_\alpha\psi$
\item $(\varphi\equiv\psi)\rightarrow(\varphi\rightarrow\psi)$
\item $(\psi\equiv\psi')\rightarrow(\varphi[x:=\psi]\equiv\varphi[x:=\psi'])$, if $x\in fvar(\varphi)$
\item $\forall x(\varphi\equiv\psi)\rightarrow (\forall x\varphi\equiv\forall x\psi)$
\item $\forall x\varphi\rightarrow\varphi[x:=\psi]$
\item $\forall x (\varphi\rightarrow\psi)\rightarrow (\forall x\varphi\rightarrow\forall x\psi)$
\item $\forall x (\varphi\rightarrow\psi)\rightarrow (\varphi\rightarrow\forall x\psi)$, if $x\notin fvar(\varphi)$
\item $\square\forall x\varphi\rightarrow\forall x\square\varphi$
\item $\forall x\square\varphi\rightarrow\square\forall x\varphi$ (Barcan formula)
\end{enumerate}

The set $\mathbb{AX}$ of all axioms is the smallest set that contains all formulas (i)--(xiii) above and is closed under the following condition (*): If $\varphi$ is an axiom and $x\in fvar(\varphi)$, then $\forall x\varphi$ is an axiom.

The rules of inference are:
\begin{itemize}
\item Modus Ponens MP: ``From $\varphi$ and $\varphi\rightarrow\psi$ infer $\psi$."
\item Axiom Necessitation AN: ``If $\varphi$ is an axiom, then infer $\square\varphi$."
\end{itemize}

The resulting deductive system is an amalgam of basic non-Fregean logic $\mathit{SCI}$ (propositional logic + the axioms of propositional identity (v)--(vii)) and Lewis modal logic $S3$ (propositional logic + axioms (ii)--(iv) + rule AN) together with axioms for propositional quantification (axioms (ix)--(xiii)) and bridge axiom (viii).\footnote{We follow a Lemmon-style axiomatization of $S3$, see, e.g., \cite{hug}, pp. 199. Note that stating axiom (viii) implies that variable $x$ occurs free in both $\varphi$ and $\psi$.} We refer to that system as $S3^\forall_\equiv$. $S4^\forall_\equiv$ is the system that results from adding the axiom scheme $\square\varphi\rightarrow\square\square\varphi$. $S5^\forall_\equiv$ is obtained by adding the scheme $\neg\square\varphi\rightarrow\square\neg\square\varphi$ to $S4^\forall_\equiv$. Since the Necessitation Rule is not part of the deductive system, we are able to define the notion of \textit{derivation} in the same natural way as in (non-modal) propositional logic: a derivation of $\varphi\in Fm(C)$ from $\Phi\subseteq Fm(C)$ is a finite sequence of formulas $\varphi_1,...,\varphi_n=\varphi$ such that for each $i=1,...,n$: $\varphi_i\in \Phi$ or $\varphi_i$ is an axiom or $\varphi_i$ is obtained by rule AN or $\varphi_i$ is obtained by rule MP applied to formulas $\varphi_j$, $\varphi_k=\varphi_j\rightarrow\varphi_i$, where $j,k < i$.

Usually, the Barcan formula (axiom (xiii)) refers to a certain semantic property of first-order modal logics and in that context it has been the object of some philosophical debates. The Barcan formula is also considered as an axiom in the approaches to propositional quantifiers presented by Fine \cite{fin} and Bull \cite{bul}. In fact, the Barcan formula as well as its converse (axiom (xii)) are valid in the possible worlds semantics. In our approach, the Barcan formula corresponds to a semantic property which is used to establish soundness of Axiom Necessitation (see the first equivalence of \eqref{10} after Definition \ref{200} below). The converse of the Barcan formula ensures that a weak Generalization Rule holds, see Lemma \ref{110} below.\footnote{In contrast to \cite{fin}, our system does not contain the full Generalization Rule.} Note that if propositional identity $\varphi\equiv\psi$ is given by strict equivalence $\square(\varphi\leftrightarrow\psi)$, then the bridge axiom (viii) is derivable from the Barcan formula. In the proof of the Completeness Theorem, axiom (viii) ensures that a certain higher-order function on the universe of the constructed model is well-defined. In the simpler and weaker semantics defined in the last section, models do not contain that higher-order function and the Barcan formula as well as axiom (viii) can be avoided. 

\begin{definition}\label{80}
If $\Phi\cup\{\varphi\}\subseteq Fm(C)$, then we write $\Phi\vdash_m\varphi$ in order to express that there is a derivation of $\varphi$ from $\Phi$ in system $Sm^\forall_\equiv$, where $m\in\{3,4,5\}$. 
\end{definition}

\begin{lemma}[Deduction Theorem]\label{100}
If $\Phi\cup\{\varphi\}\vdash_m\psi$, then $\Phi\vdash_m\varphi\rightarrow\psi$, for $m\in\{3,4,5\}$.
\end{lemma}

\paragraph*{Proof.}
It is enough to consider $m=3$. The assertion can be shown by induction on the length $n$ of a derivation of $\psi$ from $\Phi\cup\{\varphi\}$. If $n=1$, then $\psi$ is an axiom or $\psi\in\Phi\cup\{\varphi\}$ or $\psi$ is obtained by the rule of Axiom Necessitation AN. In the first two cases, the assertion follows from standard arguments using classical propositional logic. Suppose $\psi=\square\psi'$ for some axiom $\psi'$. Then $\Phi\vdash_3\psi'$. By AN, $\Phi\vdash_3\square\psi'$. Since $\square\psi'\rightarrow(\varphi\rightarrow\square\psi')$ is an axiom (a substitution-instance of a propositional tautology), MP yields the assertion. Now suppose $n>1$ and the claim is true for all derivations of length $\le n-1$. We may assume that the last step in the derivation is MP (all other cases follow in the same way as before). The assertion then follows from axioms of propositional logic. Q.E.D.

\begin{lemma}[Generalization]\label{110}
If $\Phi\vdash_m\varphi$ and $x\in fvar(\varphi)\smallsetminus fvar(\Phi)$, then $\Phi\vdash_m\forall x\varphi$, for $m\in\{3,4,5\}$.
\end{lemma}

\paragraph*{Proof.}
As before, we consider $m=3$ and show the assertion by induction on the length $n$ of a derivation. If $n=1$ and the conditions of the Lemma hold, then $\varphi$ is an axiom or it is obtained by AN (note that $\varphi\in \Phi$ is impossible). In the first case, $\forall x\varphi$ is an axiom and therefore $\Phi\vdash_3\forall x\varphi$. In the second case, $\varphi=\square\varphi'$ for some axiom $\varphi'$. Then $\forall x\varphi'$ is an axiom, and by AN we obtain $\Phi\vdash_3\square\forall x\varphi'$. Axiom (xii) and MP yield the assertion. Now we suppose $n>1$ and the assertion holds for all derivations of length $\le n-1$. We may assume that the last step of the derivation is MP. There are formulas $\psi$ and $\psi\rightarrow\varphi$ derived in less steps. If $x\in fvar(\psi)$, then by induction hypothesis: $\Phi\vdash_3\forall x\psi$ and $\Phi\vdash_3\forall x(\psi\rightarrow\varphi)$. The assertion then follows from axiom (x) and MP. Now suppose $x\notin fvar(\psi)$. Since $x\in fvar(\varphi)$, the induction hypothesis yields $\Phi\vdash_3\forall x(\psi\rightarrow\varphi)$. By axiom (xi) and MP, $\Phi\vdash_3\psi\rightarrow\forall x\varphi$. MP yields the assertion. Q.E.D.

\begin{lemma}[Necessitation]\label{120}
In $S4^\forall_\equiv$ and $S5^\forall_\equiv$, the Necessitation Principle holds. That is, for any $\varphi\in Fm(C)$, if $\vdash_m\varphi$, then $\vdash_m\square\varphi$, for $m\in\{4,5\}$.
\end{lemma}

\paragraph*{Proof.}
We fix $m=4$ and show the assertion by induction on the length $n$ of a derivation of $\varphi$ from the empty set. If $n=1$, then $\varphi$ is an axiom or $\varphi$ is derived by the rule AN. In the former case, AN yields $\vdash_4\square\varphi$. In the latter case, there is an axiom $\psi$ such that $\varphi=\square\psi$. Then the axiom $\square\psi\rightarrow\square\square\psi$ and the rule of MP yield $\vdash_4\square\varphi$. Now suppose there is a derivation of $\varphi$ of length $n>1$. We may assume that the last step is MP. There are derivations of formulas $\psi$ and $\psi\rightarrow\varphi$ of length less than $n$, respectively. By induction hypothesis, $\square\psi$ and $\square(\psi\rightarrow\varphi)$ are derivable from the empty set. Axiom (iii) and MP yield $\vdash_4\square\varphi$. Q.E.D.

\begin{lemma}\label{130}
For any $\varphi,\psi\in Fm(C)$, $\vdash_m(\varphi\equiv\psi)\rightarrow\square(\varphi\equiv\psi)$, for $m\in\{3,4,5\}$.
\end{lemma}

\paragraph*{Proof.}
It suffices to consider $m=3$. Then\\
$\vdash_3(\varphi\equiv\psi)\rightarrow ((\varphi\equiv x)[x:=\varphi]\equiv (\varphi\equiv x)[x:=\psi])$, by axiom (vii), where $x\notin fvar(\varphi)$\\
$\vdash_3((\varphi\equiv\varphi)\equiv(\varphi\equiv\psi))\rightarrow (\square x[x:=(\varphi\equiv\varphi)]\equiv\square x[x:=(\varphi\equiv\psi)])$, again by axiom (vii)\\
$\vdash_3(\varphi\equiv\psi)\rightarrow (\square(\varphi\equiv\varphi)\equiv\square(\varphi\equiv\psi))$, by transitivity of implication in propositional logic\\
$\vdash_3(\varphi\equiv\psi)\rightarrow (\square(\varphi\equiv\varphi)\rightarrow\square(\varphi\equiv\psi))$, by axiom (ii) and transitivity of implication\\
$\vdash_3(\varphi\equiv\psi)\rightarrow\square(\varphi\equiv\psi)$, since $\square(\varphi\equiv\varphi)$ is a theorem (apply AN to axiom (v))\\
Q.E.D.

\section{Denotational semantics}

Recall that a preorder is a binary relation which is reflexive and transitive (but not necessarily anti-symmetric). There are several ways to introduce Boolean prealgebras (see, e.g., \cite{fox, pol}). We propose the following definition.

\begin{definition}\label{140}
Let $\mathcal{M}=(M, f_\bot, f_\top, f_\neg, f_\vee, f_\wedge, f_\rightarrow, \le_\mathcal{M})$ be a structure with universe $M$, operations $f_\bot, f_\top, f_\neg, f_\vee, f_\wedge, f_\rightarrow$ on $M$ of type $0,0,1,2,2,2$, respectively, and a preorder $\le_\mathcal{M}$ on $M$. We call $\mathcal{M}$ a Boolean prealgebra (or a Boolean prelattice) if the equivalence relation $\approx_\mathcal{M}$ defined by
\begin{equation*}
a\approx_\mathcal{M} b:\Leftrightarrow a\le_\mathcal{M} b\text{ and }b\le_\mathcal{M} a
\end{equation*}
is a congruence relation on $M$ and the quotient algebra of $\mathcal{M}$ modulo $\approx_\mathcal{M}$ is a Boolean algebra with lattice order $\le'$ given by $\overline{a}\le'\overline{ b}\Leftrightarrow a\le_\mathcal{M} b$, and induced operations $\overline{f_\bot},\overline{f_\top}, \overline{f_\neg}, \overline{f_\vee}, \overline{f_\wedge}, \overline{f_\rightarrow}$ for bottom and top element, complement, supremum (join), infimum (meet) and implication, respectively ($\overline{a},\overline{b}$ denote the congruence classes of $a,b\in M$ modulo $\approx_\mathcal{M}$).

A filter $F$ (with respect to $\le_\mathcal{M}$) in a Boolean prealgebra $\mathcal{M}$ is a non-empty subset $F\subseteq M$ such that for all $a,b\in M$ the usual filter axioms hold:
\begin{itemize}
\item if $a\in F$ and $a\le_\mathcal{M} b$, then $b\in F$
\item if $a,b\in F$, then $f_\wedge(a,b)\in F$
\item $f_\bot\notin F$
\end{itemize}
An ultrafilter (or prime filter) w.r.t. $\le_\mathcal{M}$ is a maximal filter w.r.t. $\le_\mathcal{M}$.\footnote{Prime filters, in its general form, are defined in a different way. Recall, however, that in a Boolean lattice every prime filter is a maximal filter, i.e., both concepts coincide. In \cite{blosus1}, the ``truth-set" of a $\mathit{SCI}$-model is defined in terms of prime filters.}
\end{definition} 

Notice that any filter of a Boolean prealgebra contains the element $f_\top$ because in the quotient Boolean algebra the top element $\overline{f_\top}$ is contained in every lattice filter. 

\begin{lemma}\label{170}
Let $\mathcal{M}$ be a Boolean prealgebra with preorder $\le_\mathcal{M}$ and let $F$ be a filter w.r.t. $\le_\mathcal{M}$. The following conditions are equivalent.
\begin{enumerate}
\item $F$ is the smallest filter, i.e., the intersection of all (ultra)filters w.r.t. $\le_\mathcal{M}$.
\item $F=\{a\in M\mid a\approx_\mathcal{M} f_\top\}$.
\item $a\le_\mathcal{M} b\Leftrightarrow f_\rightarrow(a,b)\in F$, for all $a,b\in M$. 
\end{enumerate}
\end{lemma}

\paragraph*{Proof.}
(iii)$\rightarrow$(ii):  Let $a\in F$. Since $f_\rightarrow(a,f_\rightarrow(f_\top,a))$ represents a propositional tautology, it equals the top element of the quotient Boolean algebra. Hence, it is an element of any filter of the Boolean prealgebra, in particular of $F$. By (iii), $f_\top\le_\mathcal{M} a$. Also $a\le_\mathcal{M} f_\top$ because $\overline{f_\top}$ is the top element of the quotient algebra. Now (ii) follows.\\
(ii)$\rightarrow$(i): Let $G$ be any filter. If $a\in F$, then $a\approx_\mathcal{M}f_\top\in G$. Since $G$ is a filter, $a\in G$. It follows that $F\subseteq G$. Thus, $F$ is the smallest filter.\\
(i)$\rightarrow$(iii): $a\le_\mathcal{M} b$ iff $\overline{a}\le'\overline{b}$ in the quotient algebra with lattice order $\le'$ iff $\overline{f_\rightarrow}(\overline{a},\overline{b})=\overline{f_\top}$ (as in any Boolean algebra). By (i), $F$ is the smallest filter of the Boolean prealgebra. One easily shows that the canonical homomorphism $a\mapsto\overline{a}$ maps $F$ to the smallest lattice filter of the quotient algebra, i.e., to $\overline{f_\top}$. Hence, the last condition is equivalent with $f_\rightarrow(a,b)\in F$. Q.E.D.\\

If $\mathcal{M}$ is a Boolean prealgebra with preorder $\le_\mathcal{M}$, then it is possible that $\mathcal{M}$ is already a Boolean algebra and $\le_\mathcal{M}$ is not the lattice order $\le$. In this case, $\le$ refines $\le_\mathcal{M}$. For, $a\le b\Leftrightarrow f_\rightarrow(a,b)=f_\top\Rightarrow a\le_\mathcal{M} b$. Thus, the smallest filter $F$ w.r.t. $\le_\mathcal{M}$ is a lattice filter of the Boolean algebra $\mathcal{M}$, i.e., a filter w.r.t. $\le$. The quotient algebra of $\mathcal{M}$ modulo $\approx_\mathcal{M}$ (i.e., modulo the lattice filter $F$) then is a further Boolean algebra.\\

Boolean prealgebras are considered as models in research on \textit{Hyperintensions} where logical modeling is investigated mainly from the viewpoint of natural language semantics (see, e.g., \cite{fox,pol}). It is argued that possible worlds semantics does not provide enough intensions for the modeling of natural language meanings. Solutions are discussed where propositions are viewed as elements of Boolean prealgebras. However, a connection to \textit{Non-Fregean Logic}, found in the next theorem, seems to have been unnoticed so far. Boolean prealgebras and models of $\mathit{SCI}$ are essentially the same objects:

\begin{theorem}\label{190}
The following assertions (a)--(c) hold true.\\
(a) If $\mathcal{M}=(M, f_\bot, f_\top, f_\neg, f_\vee, f_\wedge, f_\rightarrow, \le_\mathcal{M})$ is a Boolean prealgebra, then $\mathcal{M}'=(M, \mathit{TRUE}, f_\bot, f_\top, f_\neg, f_\vee, f_\wedge, f_\rightarrow, f_\equiv)$ is a model of $\mathit{SCI}$, where $\mathit{TRUE}$ is an ultrafilter w.r.t. the preorder $\le_\mathcal{M}$ and $f_\equiv$ is any binary function such that $f_\equiv(a,b)\in \mathit{TRUE}\Leftrightarrow a=b$, for all $a,b\in M$.\\
(b) Suppose $\mathcal{M}=(M, \mathit{TRUE}, f_\bot, f_\top, f_\neg, f_\vee, f_\wedge, f_\rightarrow, f_\equiv)$ is a model of $\mathit{SCI}$. Let $F$ be the intersection of all sets $T\subseteq M$ such that 
\begin{equation*}
(M, T, f_\bot, f_\top, f_\neg, f_\vee, f_\wedge, f_\rightarrow, f_\equiv)
\end{equation*}
is a model of $\mathit{SCI}$. Define $a\le_\mathcal{M'}b :\Leftrightarrow f_\rightarrow(a,b)\in F$. Then $\le_\mathcal{M'}$ is a preorder on $M$ and $\mathcal{M}':= (M, f_\bot, f_\top, f_\neg, f_\vee, f_\wedge, f_\rightarrow, \le_\mathcal{M'})$ is a Boolean prealgebra such that the sets $T$ are ultrafilters and $F$ is the smallest filter w.r.t. $\le_\mathcal{M'}$.\\
(c) The transformations described in (a) and (b) are in the following sense inverse to each other. If $\mathcal{M}$ is a Boolean prealgebra, then $\mathcal{M}''=\mathcal{M}$; and if $\mathcal{M}$ is a $\mathit{SCI}$-model, then one can find an ultrafilter of $\mathcal{M}'$ and a function $f_\equiv$ such that $\mathcal{M}''=\mathcal{M}$.
\end{theorem}

\paragraph*{Proof.}
The proof of (a) is straightforward. We prove (b). One easily checks that $\le_\mathcal{M'}$ is a preorder, $F$ is a filter and all sets $T$ such as given in the theorem are ultrafilters w.r.t. $\le_\mathcal{M'}$. From the definition of $F$ it follows that $F=\{a\in M\mid a\approx_\mathcal{M'} f_\top\}$ and $\approx_\mathcal{M'}$ is a congruence relation. Then for the quotient algebra we get $\overline{a}\le'\overline{b}$ iff $\overline{f_\rightarrow}(\overline{a},\overline{b})=\overline{f_\top}$, where $\le'$ is the partial order as given in the definition and $\overline{f_\top}=F$. It follows that the quotient algebra is a Boolean algebra with lattice order $\le'$. \\
Finally, we show (c). Let $\mathcal{M}$ be a Boolean prealgebra. Then we obtain the $\mathit{SCI}$-model $\mathcal{M'}$ according to (a). From $\mathcal{M'}$ we obtain the Boolean prealgebra $\mathcal{M''}$ in accordance with the construction in (b). By Lemma \ref{170}, the preorder of $\mathcal{M}$ is exactly the preorder defined for $\mathcal{M''}$. Also the universes and operations are the same. Thus, $\mathcal{M}=\mathcal{M''}$. The second part of the assertion follows readily from the construction. Q.E.D.\\

We observe that for a given model of $\mathit{SCI}$ one may find a Boolean prealgebra in a simpler way. Suppose $\mathcal{M}=(M, \mathit{TRUE}, f_\bot, f_\top, f_\neg, f_\vee, f_\wedge, f_\rightarrow, f_\equiv)$ is a model of $\mathit{SCI}$. Define $a\le_\mathcal{M'} b:\Leftrightarrow f_\rightarrow(a,b)\in \mathit{TRUE}$. Then, $\le_\mathcal{M'}$ is a preorder and $\mathcal{M'}:=(M, f_\bot, f_\top, f_\neg, f_\vee, f_\wedge, f_\rightarrow, \le_\mathcal{M'})$ is a Boolean prealgebra. In fact, the quotient algebra modulo $\approx_\mathcal{M'}$ is the two-element Boolean algebra.

\begin{definition}\label{200}
A propositional domain for the language $Fm(C)$ is a structure
\begin{equation*}
\mathcal{M}=(M, \mathit{TRUE}, \mathit{NEC}, f_\bot, f_\top, f_\square, f_\neg, f_\vee, f_\wedge,f_\rightarrow, f_\equiv, f_\forall, \varGamma)
\end{equation*}
where $M$ is a non-empty set whose elements are called \textit{propositions}, $\mathit{TRUE}\subseteq M$ is the set of true propositions, $NEC\subseteq M$ is the set of necessary propositions, $f_\bot$, $f_\top$, $f_\square$, $f_\neg$, $f_\vee$, $f_\wedge$, $f_\rightarrow$, $f_\equiv$ are operations on $M$ of type $0,0,1, 1, 2, 2, 2,2$, respectively, $f_\forall\colon M^M\rightarrow M$ is a higher-order function, and $\varGamma\colon C\rightarrow M$ is the so-called Gamma-function satisfying $\varGamma(\bot)=f_\bot$ and $\varGamma(\top)=f_\top$. An assignment for $\mathcal{M}$ is a function $\gamma\colon V\rightarrow M$. If $\gamma\in M^V$ is an assignment, $x\in V$ and $a\in M$, then $\gamma_x^a$ is the assignment which maps $x$ to $a$ and maps variables $y\neq x$ to $\gamma(y)$. An assignment $\gamma$ extends in the following way to a unique function $\gamma\colon Fm(C)\rightarrow M$. $\gamma(c)=\varGamma(c)$ for $c\in C$, $\gamma(\square\varphi)=f_\square(\gamma(\varphi))$, $\gamma(\neg\varphi)=f_\neg(\gamma(\varphi))$, $\gamma(\varphi @ \psi)=f_@(\gamma(\varphi),\gamma(\psi))$, for $@\in\{\equiv,\vee,\wedge,\rightarrow\}$, and finally $\gamma(\forall x\varphi)=f_\forall (\lambda z.\gamma_x^z(\varphi))$, where $z$ is any new variable and $\lambda z.\gamma_x^z(\varphi)$ denotes the function $m\mapsto \gamma_x^m(\varphi)$ from $M$ to $M$.\footnote{Very similar semantics for quantifiers are given in \cite{her, blo}. Note that we cannot simply interpret the universal quantifier as an infinite meet operation or as the infimum of an arbitrary (infinite) subset. This would require a complete Boolean (pre)algebra -- a condition which is apparently too strong to establish a Completeness Theorem (see the completeness proof below). Moreover, requiring the existence of countably complete (non-principal) ultrafilters would involve questions concerning the set-theoretical foundations.} Given $\varphi\in Fm$, $x\in fvar(\varphi)$, $\gamma\in M^V$, a function $t\colon M\rightarrow M$ is said to be $(\varphi,x,\gamma)$-definable if $t(m)=\gamma_x^m(\varphi)$, for all $m\in M$. A function $t\colon M\rightarrow M$ is said to be definable if $t$ is $(\varphi,x,\gamma)$-definable for some $\varphi\in Fm(C)$, $x\in fvar(\varphi)$ and $\gamma\in M^V$. A propositional domain $\mathcal{M}$ is a $S3_\equiv^\forall$-model if the following conditions hold:
\begin{enumerate}
\item If $\mathit{NEC}\neq\varnothing$, then the relation $\le_\mathcal{M}$ on $M$ defined by 
\begin{equation*}
a\le_\mathcal{M} b:\Leftrightarrow f_\rightarrow(a,b)\in \mathit{NEC} 
\end{equation*}
is a preorder and $(M, f_\bot, f_\top, f_\neg, f_\vee, f_\wedge, f_\rightarrow, \le_\mathcal{M})$ is a Boolean prelattice.
\item The following truth conditions hold for all $a,b\in M$ (even if $\mathit{NEC}=\varnothing$):
\begin{enumerate}
\item $f_\bot\in M\smallsetminus \mathit{TRUE}$, $f_\top\in \mathit{TRUE}$
\item $f_\rightarrow(a,b)\in \mathit{TRUE}\Leftrightarrow a\notin \mathit{TRUE}\text{ or }b\in \mathit{TRUE}$
\item $f_\neg(a)\in \mathit{TRUE}\Leftrightarrow a\notin \mathit{TRUE}$
\item $f_\wedge(a,b)\in \mathit{TRUE}\Leftrightarrow a\in \mathit{TRUE}$ and $b\in \mathit{TRUE}$
\item $f_\vee(a,b)\in \mathit{TRUE}\Leftrightarrow a\in \mathit{TRUE}$ or $b\in \mathit{TRUE}$
\item $f_\square(a)\in \mathit{TRUE}\Leftrightarrow a\in \mathit{NEC}$
\item $f_\equiv(a,b)\in TRUE\Leftrightarrow a=b$
\item $f_\forall(t)\in TRUE$ whenever $t\colon M\rightarrow M$ is a definable function with image $im(t)\subseteq \mathit{TRUE}$\footnote{The implication $f_\forall(t)\in \mathit{TRUE}\Rightarrow im(t)\subseteq \mathit{TRUE}$, for any definable $t$, will follow from (iv)(h) and the fact that $\mathit{TRUE}$ is a filter on $M$.}
\end{enumerate}
\item If $\mathit{NEC}\neq\varnothing$, then $\mathit{NEC}\subseteq \mathit{TRUE}$ is a filter on $M$, i.e., for all $a,b\in M$:
\begin{enumerate}
\item if $a\in \mathit{NEC}$ and $a\le_\mathcal{M} b$, then $b\in \mathit{NEC}$
\item if $a,b\in \mathit{NEC}$, then $f_\wedge(a,b)\in \mathit{NEC}$
\end{enumerate}
\item If $\mathit{NEC}\neq\varnothing$, then the following hold for all $a,b\in M$:
\begin{enumerate}
\item $f_\top\le_\mathcal{M} f_\equiv(a,a)$
\item $f_\equiv(a,b)\le_\mathcal{M} f_\rightarrow(a,b)$
\item $f_\equiv(a,b)\le_\mathcal{M} f_\equiv(t(a),t(b))$, for any definable function $t\colon M\rightarrow M$
\item $f_\square(a)\le_\mathcal{M} a$
\item $f_\square (f_\rightarrow(a,b))\le_\mathcal{M} f_\rightarrow(f_\square(a), f_\square(b))$\footnote{This condition follows from (f) and (d).}
\item $f_\square (f_\rightarrow(a,b))\le_\mathcal{M} f_\square(f_\rightarrow(f_\square(a), f_\square(b)))$
\item $f_\forall(t)\le_\mathcal{M} f_\equiv(f_\forall (t_1),f_\forall(t_2))$, whenever $t_1$ is $(\varphi,x,\gamma)$-definable, $t_2$ is $(\psi,x,\gamma)$-definable, and $t$ is the $(\varphi\equiv\psi,x,\gamma)$-definable function $t(a)= f_\equiv(t_1(a),t_2(a))$
\item $f_\forall(t)\le_\mathcal{M} t(a)$, for any definable function $t\colon M\rightarrow M$
\item $f_\forall(t)\le_\mathcal{M} f_\rightarrow (f_\forall (t_1),f_\forall(t_2))$, whenever $t_1$ is $(\varphi,x,\gamma)$-definable, $t_2$ is $(\psi,x,\gamma)$-definable, and $t$ is the $(\varphi\rightarrow\psi,x,\gamma)$-definable function $t(a)= f_\rightarrow(t_1(a),t_2(a))$
\item $f_\forall(t)\le_\mathcal{M} f_\rightarrow (a,f_\forall(t'))$, whenever $t'$ is $(\psi,x,\gamma)$-definable, and $t$ is the $(\varphi\rightarrow\psi,x,\gamma)$-definable function $t(a)= f_\rightarrow(b,t'(a))$, where $b$ is the denotation of $\varphi$ and $fvar(\varphi)=\varnothing$\footnote{The denotation of a sentence is independent of any assignment.}
\item $f_\square (f_\forall(t))\approx_\mathcal{M} f_\forall(t')$, for every definable function $t\colon M\rightarrow M$ and function $t'\colon M\rightarrow M$ with $t'(a)=f_\square(t(a))$.
\item $f_\forall(t)\in NEC$ whenever $t\colon M\rightarrow M$ is a definable function with image $im(t)\subseteq NEC$\footnote{This condition follows from (iv)(k) together with (ii)(f) and (ii)(h).}
\end{enumerate}
\end{enumerate}
A $S3_\equiv^\forall$-model is called normal if $NEC\neq\varnothing$, otherwise the model is called non-normal. A normal $S3_\equiv^\forall$-model is a $S4_\equiv^\forall$-model if $f_\square(a)\le_\mathcal{M} f_\square(f_\square(a))$ for every $a\in M$. A $S4_\equiv^\forall$-model is a $S5_\equiv^\forall$-model if $f_\neg(f_\square(a))\le_\mathcal{M} f_\square(f_\neg(f_\square(a)))$ for every $a\in M$.
\end{definition} 

Note that if $\mathit{NEC}\neq\varnothing$, then $\mathit{TRUE}$ is an ultrafilter. In order to see this, suppose $a\in\mathit{TRUE}$ and $a\le_\mathcal{M} b$. The latter condition implies $f_\rightarrow(a,b)\in\mathit{NEC}\subseteq\mathit{TRUE}$, by (i). Then by condition (ii)(b), $b\in\mathit{TRUE}$. By (ii)(a), $f_\bot\notin \mathit{TRUE}$. Together with (ii)(b), this establishes the filter conditions. Using (ii)(c) one shows that $\mathit{TRUE}$ is a maximal filter. 

Observe that the higher-order function $f_\forall\colon M^M\rightarrow M$ satisfies for every definable function $t\in M^M$ the following conditions:
\begin{equation}\label{10}
\begin{split}
&f_\forall(t)\in \mathit{NEC}\Leftrightarrow im(t)\subseteq \mathit{NEC}\\
&f_\forall(t)\in TRUE\Leftrightarrow im(t)\subseteq \mathit{TRUE}
\end{split}
\end{equation}

The first equivalence is given by the conditions (iv)(l)+(iv)(h). This equivalence is important for the soundness of rule AN: if $\varphi$ is an axiom and $x\in fvar(\varphi)$, then $\forall x\varphi$ is an axiom and, by rule AN, should be mapped to a necessary proposition. The second equivalence is given by the conditions (ii)(h)+(iv)(h) which ensure the following for any assignment $\gamma\in M^V$: $\gamma(\forall x\varphi)\in \mathit{TRUE}$ iff $\gamma_x^m(\varphi)\in\mathit{TRUE}$ for all $m\in M$. Since $\mathit{TRUE}$ and $\mathit{NEC}$ are filters, $a\approx_\mathcal{M} b$ implies ($a\in TRUE\Leftrightarrow b\in TRUE$) and ($a\in \mathit{NEC}\Leftrightarrow b\in \mathit{NEC}$). One also verifies that $\approx_\mathcal{M}$ is, by condition (iv)(f), a congruence relation with respect to $f_\square$. In fact, (iv)(f) establishes monotonicity of $f_\square$: if $a\le_\mathcal{M} b$, then $f_\square(a)\le_\mathcal{M}f_\square(b)$. However, $\approx_\mathcal{M}$ is, in general, not a congruence relation with respect to the operation $f_\equiv$. That is, $a\approx_\mathcal{M} b$ and $a'\approx_\mathcal{M} b'$ does not imply $f_\equiv(a,a')\approx_\mathcal{M} f_\equiv(b,b')$. In fact, if $a=a'$ and $b\neq b'$, then we obtain propositions $f_\equiv(a,a')\in \mathit{TRUE}$ and $f_\equiv(b,b')\notin \mathit{TRUE}$ with different truth values.

Note that for a non-normal model, the conditions (i), (iii) and (iv) are irrelevant. 

\begin{lemma}[Coincidence Lemma]\label{300}
Let $\mathcal{M}$ be a model, $\varphi\in Fm(C)$, and let $\gamma$, $\gamma'\colon V\rightarrow M$ be assignments such that $\gamma(x)=\gamma'(x)$ for all $x\in fvar(\varphi)$. Then $\gamma(\varphi)=\gamma'(\varphi)$.
\end{lemma}

The proof of the Coincidence Lemma is an induction on $\varphi$, simultaneously for all assignments $\gamma,\gamma'$. The lemma says in particular that the denotation of a sentence, i.e., a formula with no free variables, is independent of any assignment and depends only from the Gamma-function.

Observe that if $x,y$ are distinct variables, then $(\gamma_x^a)_y^b=(\gamma_y^b)_x^a$ for any assignment $\gamma$ and elements $a,b$ of the model-theoretic universe. If $x_1,...,x_n$ are pairwise distinct variables, then we write $\gamma_{x_1,...,x_n}^{a_1,...,a_2}$ for the assignment $(...((\gamma_{x_1}^{a_1})_{x_1}^{a_2})...)_{x_n}^{a_n}$.

\begin{definition}\label{310}
Let $\mathcal{M}$ be a model, $\gamma\colon V\rightarrow M$ an assignment and $\sigma\colon V\rightarrow Fm(C)$ a substitution. Then we denote the assignment $x\mapsto\gamma(\sigma(x))$ by $\gamma\sigma$.
\end{definition}

The next result is an analogue of the Substitution Lemma of classical first-order logic.

\begin{lemma}[Substitution Lemma]\label{320}
Let $\mathcal{M}$ be a model, $\gamma\colon V\rightarrow M$ an assignment and $\sigma\colon V\rightarrow Fm(C)$ a substitution. Then 
\begin{equation*}
\gamma\sigma(\varphi)=\gamma(\varphi[\sigma]).
\end{equation*}
\end{lemma}

\paragraph*{Proof.}
Induction on $\varphi$ simultaneously for all assignments $\gamma$ and all substitutions $\sigma$. The basis cases $\varphi=x$ and $\varphi=c$ follow immediately from the definition. Most of the cases of the induction step follow straightforwardly. We show the quantifier case.
Let $u\in V$ such that $u\notin fvar(\sigma(x))$ for all $x\in fvar(\forall y\psi)$. Then one easily checks that $(\gamma\sigma)_y^a(v)=\gamma_u^a\sigma[y:=u](v)$ for every $v\in fvar(\psi)$ and every $a\in M$. In the following, let $u$ be the variable forced by the substitution $\sigma$ w.r.t. $\forall y\psi$. Then:
\begin{equation*}
\begin{split}
\gamma\sigma(\forall y\psi)&=f_\forall (\lambda z.(\gamma\sigma)_y^z (\psi))\\
&=f_\forall(\lambda z.((\gamma_u^z \sigma[y:=u])(\psi))\text{ by the Coincidence Lemma}\\
&=f_\forall(\lambda z.(\gamma_u^z(\psi[\sigma[y:=u])))\text{ by the induction hypothesis}\\
&=\gamma(\forall u (\psi[\sigma[y:=u]))\\
&=\gamma((\forall y\psi)[\sigma])
\end{split}
\end{equation*}
Q.E.D.\\

Notice that the Substitution Lemma implies equations of the following form:

\begin{equation*}
\gamma_{x_1,...,x_n}^{\gamma(\varphi_1),...,\gamma(\varphi_n)}(\varphi)=\gamma(\varphi[x_1:=\varphi_1,...,x_n:=\varphi_n]).
\end{equation*}

\begin{definition}\label{360}
Let $\mathcal{M}$ be a $S3_\equiv^\forall$-model, $\gamma\colon V\rightarrow M$ an assignment and $\varphi\in Fm(C)$. Satisfaction (truth) of $\varphi$ in the interpretation $(\mathcal{M},\gamma)$ is defined as follows:
\begin{equation*}
(\mathcal{M},\gamma)\vDash\varphi :\Leftrightarrow\gamma(\varphi)\in \mathit{TRUE}.
\end{equation*}
This notion extends in the usual way to sets of formulas. For $\Phi\subseteq Fm(C)$ define $Mod_3(\Phi):=\{(\mathcal{M},\gamma)\mid\mathcal{M}$ a normal $S3_\equiv^\forall$-model, $\gamma\in M^V$ and $(\mathcal{M},\gamma)\vDash\Phi\}$. Logical consequence is defined as follows:
\begin{equation*}
\Phi\Vdash_3\varphi:\Leftrightarrow Mod_3(\Phi)\subseteq Mod_3(\{\varphi\}).
\end{equation*}
As usual, we write $\Vdash_3\varphi$ instead of $\varnothing\Vdash_3\varphi$. Logical consequence for the logics generated by the class of all normal $S4_\equiv^\forall$-models, the class of all normal $S5_\equiv^\forall$-models, respectively, are defined analogously.
\end{definition}

Note that we have defined logical consequence only with respect to the class of \textit{normal} models. This is in accordance with the situation in modal logic $S3$ where validity of a formula $\varphi$ is defined as truth of $\varphi$ in all \textit{normal} worlds in all Kripke models. 

It is not hard to show that a normal model satisfies all axioms and rules of inference. For instance, let $\varphi'$ be a substitution-instance of the propositional tautology $\varphi$. In each Boolean algebra, $\varphi$ is mapped by any assignment to the top element. Then in our Boolean prealgebras, $\varphi$ is mapped by any assignment to an element of the smallest filter containing $f_\top$ (if the model is normal, that filter is $\mathit{NEC}$) and thus to an element of $\mathit{TRUE}$. By the Substitution Lemma, the same holds for $\varphi'$. Consider now the axiom $\forall x\varphi\rightarrow\varphi[x:=\psi]$. Let $\mathcal{M}$ be a model and suppose $(\mathcal{M},\gamma)\vDash\forall x\varphi$ for some assignment $\gamma\in M^V$. Then $f_\forall(\lambda z.\gamma_x^z(\varphi))\in \mathit{TRUE}$. In particular, $\gamma_x^a(\varphi)\in \mathit{TRUE}$ where $a=\gamma(\psi)$. By the Substitution Lemma, $\gamma(\varphi[x:=\psi])=\gamma_x^a(\varphi)\in \mathit{TRUE}$. Now we consider axiom (v), $\varphi\equiv\psi$ whenever $\varphi=_\alpha\psi$. Suppose $\varphi=_\alpha\psi$. By Lemma \ref{50}, this is equivalent with the condition $\varphi[\varepsilon]=\psi[\varepsilon]$, where $\varepsilon$ is the identity substitution. We have $\gamma=\gamma\varepsilon$, for any assignment $\gamma\colon V\rightarrow M$. The Substitution Lemma implies $\gamma(\varphi)=\gamma\varepsilon(\varphi)=\gamma(\varphi[\varepsilon])=\gamma(\psi[\varepsilon])=\gamma\varepsilon(\psi)=\gamma(\psi)$. Thus, $\gamma(\varphi\equiv\psi)=f_\equiv(\gamma(\varphi),\gamma(\psi))\in \mathit{TRUE}$ and $(\mathcal{M},\gamma)\vDash\varphi\equiv\psi$. Also the soundness of axiom (vii) follows from the Substitution Lemma and the Coincidence Lemma (alternatively, one may carry out an induction on $\varphi$). We leave the remaining cases to the reader.

\begin{theorem}[Soundness]
$\Phi\vdash_m\varphi\Rightarrow\Phi\Vdash_m\varphi$, for $m\in\{3,4,5\}$.
\end{theorem}

\section{Completeness}

Completeness theorems for logics with an identity connective and quantifiers that range over a universe of denotations of formulas or sentences have been proved by several authors ( see, e.g., \cite{her, blo, str, zei}). We apply the typical Henkin construction.

\begin{lemma}\label{485}
If $\varphi$ is an axiom, $c$ a constant and $y\in V\smallsetminus var(\varphi)$, then $\varphi[c:=y]$ is an axiom. 
\end{lemma}

\paragraph*{Proof.}
The assertion is obviously true for most of the axioms. We show the assertion for axiom scheme (ix): $\forall x\varphi\rightarrow\varphi[x:=\psi]$. We have 
\begin{equation*}
\begin{split}
&(\forall x\varphi\rightarrow\varphi[x:=\psi])[c:=y]\\
&=(\forall x\varphi)[c:=y]\rightarrow(\varphi[x:=\psi])[c:=y]\\
&= \forall z(\varphi[c:=y,x:=z])\rightarrow\varphi[c:=y] [x:=\psi']\\
&=\forall z\varphi[c:=y] [x:=z]\rightarrow \varphi[c:=y] [x:=z][z:=\psi']\\
&=\forall z\chi\rightarrow \chi[z:=\psi'],
\end{split}
\end{equation*}
where $z$ is the variable forced by $[c:=y]$ w.r.t. $\forall x\varphi\rightarrow\varphi[x:=\psi]$, $\psi'=\psi[c:=y]$, and $\chi=\varphi[c:=y][x:=z]$. Note that $y\neq x$ since $y\notin var(\varphi)$. The formula $\forall z\chi\rightarrow \chi[z:=\psi']$ is clearly an axiom of scheme (ix). Q.E.D.\\

If we want to make explicit that a derivation of a formula $\varphi$ from a set $\Phi$ contains only formulas with constants from $C$, then we write $\Phi\vdash_3^C\varphi$. For a set $\Phi$ of formulas let $\Phi[c:=y]:=\{\psi[c:=y]\mid \psi\in\Phi\}$. 

\begin{lemma}[Elimination of constants]\label{490}
Let $C$ be a set of constants and let $c$ be any constant, possibly $c\notin C$. Put $C':=C\cup\{c\}$. Then $\Phi\vdash_3^{C'}\varphi$ implies $\Phi[c:=y]\vdash_3^C\varphi[c:=y]$, for almost all $y\in V$.\footnote{``for almost all $y\in V$" means \textit{for all but finitely many variables}. That is, there are only finitely many variables $y$ such that the property stated in the Lemma does not hold.} 
\end{lemma}

\paragraph*{Proof.}
We show the assertion by induction on the length $n$ of a derivation of $\varphi$ from $\Phi$ in language $Fm(C')$. If $n=1$, then $\varphi$ is an axiom or $\varphi\in\Phi$ or $\varphi$ is obtained by rule AN. By Lemma \ref{485}, if $\varphi$ is an axiom, then $\varphi[c:=y]$ is an axiom for any $y\in V\smallsetminus var(\varphi)$. It follows that in all three cases $\Phi[c:=y]\vdash_3^C\varphi[c:=y]$, if we choose $y\in V\smallsetminus var(\varphi)$. Now suppose the derivation has length $n>1$. We may assume that the last step of the derivation is Modus Ponens. Then there are formulas $\psi$, $\psi\rightarrow\varphi$ derived in less steps. By induction hypothesis, $\Phi[c:=u]\vdash_3^C\psi[c:=u]$ for almost all $u\in V$, and $\Phi[c:=z]\vdash_3^C(\psi\rightarrow\varphi)[c:=z]$ for almost all $z\in V$. But then holds both, $\Phi[c:=y]\vdash_3^C\psi[c:=y]$ and $\Phi[c:=y]\vdash_3^C(\psi\rightarrow\varphi)[c:=y]$ for almost all $y\in V$. The last formula equals $\psi[c:=y]\rightarrow\varphi[c:=y]$. MP yields the assertion. Q.E.D. 

\begin{corollary}\label{495}
Suppose $\Phi\cup\{\varphi\}\subseteq Fm(C)$, $x\in fvar(\varphi)$ and $c$ is a constant such that $c\notin con(\Phi\cup\{\varphi\})$. Then $\Phi\vdash_3\varphi[x:=c]$ implies $\Phi\vdash_3\forall x\varphi$.
\end{corollary}

\paragraph*{Proof.}
Suppose $\Phi\vdash_3\varphi[x:=c]$ and the conditions of the Corollary are satisfied. Since derivation is finitary, we may assume that $\Phi$ is a finite set. Then $var(\Phi\cup\{\varphi\})$ is finite, too. By Lemma \ref{490}, we may find an $y\in V\smallsetminus var(\Phi\cup\{\varphi\})$ such that $\Phi[c:=y]\vdash_3\varphi[x:=c][c:=y]$. Hence, $\Phi\vdash_3\varphi[x:=y]$ ($c$ does not occur in $\Phi\cup\{\varphi\}$). Because $y$ does not occur (free) in $\Phi$, we may apply Lemma \ref{110} which yields $\Phi\vdash_3\forall y (\varphi[x:=y])$. This formula is alpha-congruent with $\forall x\varphi$. Then the axioms (v) and (vi) together with MP yield $\Phi\vdash_3\forall x\varphi$. Q.E.D.\\

In our treatment of Henkin sets (Definitions \ref{520} and \ref{540}, Lemma \ref{560}) we adopt some ideas and notations from \cite{rau}. 

\begin{definition}\label{520}
A set $\Phi\subseteq Fm(C)$ is called a Henkin set if 
\begin{itemize}
\item $\Phi$ is maximally consistent
\item $\Phi\vdash_3\forall x\varphi \Leftrightarrow \Phi\vdash_3\varphi[x:=c]$ for all $c\in C$
\end{itemize}
\end{definition}

The next observation follows immediately from axioms (xii) and (xiii).

\begin{lemma}\label{530}
Let $\Phi\subseteq Fm(C)$ be a Henkin set. Then:
\begin{equation*}
\Phi\vdash_3\square\forall x\varphi\Leftrightarrow\Phi\vdash_3\square\varphi[x:=c]\text{ for all }c\in C.
\end{equation*}
\end{lemma}

\begin{definition}\label{540}
To each pair $\varphi,x$, where $\varphi\in Fm(C)$ and $x\in fvar(\varphi)$, we assign exactly one new constant $c_{\varphi,x}\notin C$ and define 
\begin{equation*}
\varphi^x:=\neg(\neg \forall x\varphi\rightarrow\neg\varphi[x:=c_{\varphi,x}]).
\end{equation*}
Furthermore, $Y(C):=\{\neg(\varphi^x)\mid \varphi\in Fm(C), x\in fvar(\varphi)\}$.
\end{definition} 

Note that $\neg(\varphi^x)$ can be written as $\exists x\neg\varphi\rightarrow\neg\varphi[x:=c_{\varphi,x}]$. In this sense, $c_{\varphi,x}$ can be seen as a witness for the truth of $\exists x\neg\varphi$.

\begin{lemma}\label{560}
If $\Phi\subseteq Fm(C)$ is consistent, then so is $\Phi\cup Y(C)\subseteq Fm(C')$, where $C'=C\cup\{c_{\varphi,x}\mid\varphi\in Fm(C), x\in V\}$ according to Definition \ref{540}.
\end{lemma}

\paragraph*{Proof.}
Suppose $\Phi\cup Y(C)\subseteq Fm(C')$ is inconsistent. There are formulas $\neg(\varphi_0^{x_0})$, ...,$\neg(\varphi_n^{x_n})\in Y(C)$ such that $\Phi\cup\{\neg(\varphi_i^{x_i})\mid i\le n\}$ is inconsistent. We may assume that $n$ is minimal with this property. Let $x:=x_n$, $\varphi:=\varphi_n$, $c:=c_{n,\varphi}$, $\Phi':=\Phi\cup\{\neg(\varphi_i^{x_i})\mid i< n\}$. Then $\Phi'$ is consistent and $\Phi'\cup\{\neg(\varphi^x)\}$ is inconsistent. In particular, $\Phi'\cup\{\neg(\varphi^x)\}\vdash_3\bot$. By the Deduction Theorem, $\Phi'\vdash_3\neg(\varphi^x)\rightarrow\bot$. Contra-position yields $\Phi'\vdash_3\top\rightarrow\varphi^x$. By MP, $\Phi'\vdash_3\neg(\neg \forall x\varphi\rightarrow\neg\varphi[x:=c])$. This yields $\Phi'\vdash_3\neg\forall x\varphi$ and $\Phi'\vdash_3\varphi[x:=c]$. By construction, $c\notin con(\varphi)\cup con(\Phi')$. We may apply Corollary \ref{495} and obtain $\Phi'\vdash_3\forall x\varphi$ and $\Phi'\vdash_3\neg\forall x\varphi$. But then $\Phi'$ is inconsistent, a contradiction. Hence, $\Phi\cup Y(C)\subseteq Fm(C')$ is consistent. Q.E.D.

\begin{definition}\label{580}
Let $\Phi\subseteq Fm(C)$ be maximally consistent. For $\varphi,\psi\in Fm(C)$ define $\varphi\approx_\Phi\psi :\Leftrightarrow\Phi\vdash_3\varphi\equiv\psi$.
\end{definition}

\begin{lemma}\label{585}
Let $\Phi\subseteq Fm(C)$ be maximally consistent. Then $\approx_\Phi$ is an equivalence relation on $Fm(C)$ containing alpha-congruence and satisfying the following:
if $\varphi_1\approx_\Phi\psi_1$ and $\varphi_2\approx_\Phi\psi_2$, then $\neg\varphi_1\approx_\Phi\neg\psi_1$, $\square\varphi_1\approx_\Phi\square\psi_1$, $\varphi_1 @\varphi_2\approx_\Phi\psi_1 @\psi_2$, where $@\in\{\vee,\wedge,\rightarrow,\equiv\}$. That is, $\approx_\Phi$ is a congruence relation on $Fm(C)$ containing alpha-congruence.
\end{lemma}

\paragraph*{Proof.}
By axiom (v), $\approx_\Phi$ is reflexive and contains alpha-congruence. Suppose $\varphi\approx_\Phi\psi$ and consider the formula $x\equiv\varphi$, where $x\in V\smallsetminus var(\varphi)$. Since $\varphi\equiv\varphi$ is an axiom, the axiom $(\varphi\equiv\psi)\rightarrow ((x\equiv\varphi)[x:=\varphi]\equiv(x\equiv\varphi)[x:=\psi])$ together with MP yields $\psi\approx_\Phi\varphi$. Thus, the relation is symmetric. Now let $\varphi\approx_\Phi\psi$ and $\psi\approx_\Phi\chi$. Let $\delta:=(x\equiv\chi)$, where $x\in V\smallsetminus var(\chi)$. By axiom (vii) and MP, $\delta[x:=\varphi]\approx_\Phi\delta[x:=\psi])$. By hypothesis, $\Phi\vdash_3\delta[x:=\psi]$. Symmetry of $\approx_\Phi$, axiom (vi) and MP yield $\Phi\vdash_3\delta[x:=\varphi]$. That is, $\varphi\approx_\Phi\chi$ and $\approx_\Phi$ is transitive. Now suppose $\varphi_1\approx_\Phi\psi_1$ and $\varphi_2\approx_\Phi\psi_2$. Let $x\in V\smallsetminus var(\psi_2)$ and $y\in V\smallsetminus var(\varphi_1)$. By axiom (vii), $(\varphi_1\wedge\varphi_2)=(\varphi_1\wedge y)[y:=\varphi_2]\approx_\Phi (\varphi_1\wedge y)[y:=\psi_2]=(\varphi_1\wedge \psi_2)=(x\wedge \psi_2)[x:=\varphi_1]\approx_\Phi (x\wedge\psi_2)[x:=\psi_1]=\psi_1\wedge\psi_2$. The remaining cases follow in a similar way. Q.E.D.\\

Propositional logic, axiom (vi) and symmetry of $\approx_\Phi$ imply the next result.

\begin{lemma}\label{595}
Let $\Phi$ be maximally consistent and $\varphi,\psi\in Fm(C)$. Then:
\begin{itemize} 
\item $\varphi\in\Phi$ iff $\Phi\vdash_3\varphi$.
\item If $\varphi\approx_\Phi\psi$, then $\varphi\in\Phi\Leftrightarrow\psi\in \Phi$.
\end{itemize}
\end{lemma}

\begin{theorem}\label{620}
Every Henkin set has a normal model.
\end{theorem}

\paragraph*{Proof.}
Let $\Phi\subseteq Fm(C)$ be a Henkin set. By $\overline{\varphi}$ we denote the equivalence class of $\varphi\in Fm(C)$ modulo $\approx_\Phi$.\\ 
\textbf{Claim} 1: For every $\varphi\in Fm(C)$ there is a $c\in C$ such that $c\approx_\Phi\varphi$.\\
\textit{Proof of the Claim}: If $x\in V\smallsetminus var(\varphi)$, then obviously $\Phi\vdash_3 (x\equiv \varphi)[x:=\varphi]$. Contra-position of axiom (ix) yields: $\Phi\vdash_3 (x\equiv \varphi)[x:=\varphi]\rightarrow\neg\forall x\neg (x\equiv\varphi)$. By MP: $\Phi\vdash_3\neg\forall x\neg (x\equiv\varphi)$. Since $\Phi$ is consistent, $\Phi\nvdash_3\forall x\neg (x\equiv \varphi)$. Because $\Phi$ is a Henkin set, $\Phi\nvdash_3\neg (c\equiv \varphi)$ for some $c\in C$. $\Phi$ is maximally consistent, thus $\Phi\vdash_3 c\equiv \varphi$. This proves Claim 1. Our model $\mathcal{M}$ is given by the following:
\begin{equation*}
\begin{split}
&M:=\{\overline{\varphi}\mid \varphi\in Fm(C)\}\\
&\mathit{TRUE}:=\{\overline{\varphi}\mid \varphi\in\Phi\}\\ 
&\mathit{NEC}:=\{\overline{\varphi}\mid\square\varphi\in\Phi\}\\
&f_\top:=\overline{\top}, f_\bot:=\overline{\bot}, f_\square(\overline{\varphi}):=\overline{\square\varphi}, f_\neg(\overline{\varphi}):=\overline{\neg\varphi}\\
&f_\rightarrow(\overline{\varphi},\overline{\psi}):=\overline{\varphi\rightarrow\psi}, f_\equiv(\overline{\varphi},\overline{\psi}):=\overline{\varphi\equiv\psi}\\ 
&f_\vee(\overline{\varphi},\overline{\psi}):=\overline{\varphi\vee\psi}, f_\wedge(\overline{\varphi},\overline{\psi}):=\overline{\varphi\wedge\psi}\\
&\varGamma(c):=\overline{c}
\end{split}
\end{equation*}
By the previous results, all these ingredients are well-defined. Furthermore, for $t\in M^M$ we define 
\begin{equation*}
\begin{split}
f_\forall(t):=
\begin{cases}
&\overline{\forall x\varphi},\text{ if there is a }\varphi\text{ such that }t(\overline{c})=\overline{\varphi[x:=c]}\text{ for all }c\in C\\
&f_\top,\text{ if such a formula }\varphi\text{ does not exist}
\end{cases}
\end{split}
\end{equation*}
Note that Claim 1 implies $M=\{\overline{c}\mid c\in C\}$. It remains to show that $f_\forall$ is well-defined. Let $t\in M^M$ and suppose there are two formulas $\varphi,\psi\in Fm(C)$ such that $\overline{\varphi[x:=c]}=t(\overline{c})=\overline{\psi[y:=c]}$ for all $c\in C$. Without lost of generality, we may assume that $x\notin var(\psi)$. Then $\varphi[x:=c]\approx_\Phi \psi[y:=c]= (\psi[y:=x])[x:=c]$, for all $c\in C$. Since $\Phi$ is a Henkin set, $\Phi\vdash_3\forall x(\varphi\equiv (\psi[y:=x]))$. By axiom (viii), $\Phi\vdash_3\forall x\varphi\equiv\forall x (\psi[y:=x])$. Note that $\forall x(\psi[y:=x])=_\alpha\forall y\psi$. By axiom (v) and transitivity of $\approx_\Phi$ we get $\forall x\varphi\approx_\Phi\forall y\psi$, that is, $\overline{\forall x\varphi}=\overline{\forall y\psi}=f_\forall(t)$. Thus, $f_\forall$ is well-defined. For each $\overline{\forall x\varphi}\in M$, the function $t\in M^M$, given by $t(\overline{c})=\overline{\varphi[x:=c]}$, is definable in the sense of Definition \ref{200}. This follows from the proof of Claim 2 below. Now it is not difficult to verify that $\mathcal{M}$ is a normal $S3^\forall_\equiv$-model. In particular, all truth conditions are satisfied. We only consider the conditions (ii)(g) and (iv)(a). We have $\overline{\varphi}=\overline{\psi}$ iff $\varphi\equiv\psi\in \Phi$ iff $f_\equiv(\overline{\varphi},\overline{\psi})=\overline{\varphi\equiv\psi}\in \mathit{TRUE}$. This shows condition (ii)(g). Furthermore, if $\overline{\varphi}=\overline{\psi}$, then $\varphi\equiv\psi\in\Phi$. By Lemma \ref{130} and MP, $\square(\varphi\equiv\psi)\in \Phi$. Hence, $f_\equiv(\overline{\varphi},\overline{\psi})=\overline{\varphi\equiv\psi}\in \mathit{NEC}$. Thus, condition (iv)(a) holds. Now let $\beta\colon V\rightarrow M$ be the assignment defined by $x\rightarrow\overline{x}$. We show that the interpretation $(\mathcal{M},\beta)$ is a model of $\Phi$. \\
\textbf{Claim} 2: $\beta(\varphi)=\overline{\varphi}$, for all $\varphi\in Fm(C)$. \\
\textit{Proof of the Claim}: Induction on the quantifier rank $qr(\varphi)$ of $\varphi$. By induction on the construction of quantifier-free formulas one easily shows that the assertion is true for all formulas of quantifier rank $0$. Now suppose the assertion is true for all formulas of quantifier rank $n$. Let $qr(\psi)=n$ and $\varphi=\forall x\psi$. Then $\beta(\varphi)=\beta(\forall x\psi)=f_\forall (\lambda z\beta_x^z(\psi))$. Consider the function $t$ defined by $t(\overline{c}):=\beta_x^{\overline{c}}(\psi)$. Then $t(z)=\lambda z\beta_x^z(\psi)$. The Substitution Lemma and the induction hypothesis yield: $t(\overline{c})=\beta_x^{\overline{c}}(\psi)=\beta(\psi[x:=c])=\overline{\psi[x:=c]}$ for all $c\in C$ (note that $qr(\psi[x:=c])<qr(\forall x\psi)$). Hence, $\beta(\forall x\psi)=f_\forall(t)=\overline{\forall x\psi}$. Hence, $\beta(\forall x\psi)=f_\forall(t)=\overline{\forall x\psi}$. So the Claim is true. Consequently: 
\begin{equation*}
(\mathcal{M},\beta)\vDash\varphi\Leftrightarrow\beta(\varphi)=\overline{\varphi}\in \mathit{TRUE}\Leftrightarrow\varphi\in\Phi.
\end{equation*}
Q.E.D.

\begin{theorem}\label{640}
Every consistent set has a normal model.
\end{theorem}

\paragraph*{Proof.}
Let $\Phi\subseteq Fm(C)$ be consistent. We extend $\Phi$ to a Henkin set $\Phi^*$ in an extended language $Fm(C^*)$, $C\subseteq C^*$. Theorem \ref{620} guarantees the existence of a normal model of $\Phi^*$. Its \textit{reduct} w.r.t. the sublanguage $Fm(C)$ then will be the desired model of $\Phi$. Let $C_0:=C$, $\Phi_0:=\Phi$. If $C_n$ and $\Phi_n\subseteq Fm(C_n)$ are already defined, then define 
\begin{equation*}
\begin{split}
&C_{n+1}:=C_n\cup\{c_{\varphi,x}\mid\varphi\in Fm(C_n), x\in fvar(\varphi)\}\\
&\Phi_{n+1}:=\Phi_n\cup Y(C_n)
\end{split}
\end{equation*}
according to the notation of Definition \ref{540}. By Lemma \ref{560}, $\Phi_{n+1}$ is consistent in $Fm(C_{n+1})$. Finally, we put $\Phi^+:=\bigcup_{n<\omega}\Phi_n$. It follows that $\Phi^+\subseteq Fm(C^*)$, where $C^*=\bigcup_{n<\omega}C_n$. Since derivation is finitary, $\Phi^+$ is consistent in the language $Fm(C^*)$. By a standard argument based on Zorn's Lemma, $\Phi^+$ extends to a maximally consistent set $\Phi^*\subseteq Fm(C^*)$. If $\Phi^*\vdash_3\forall x\varphi$, then by axiom (ix): $\Phi^*\vdash_3\varphi[x:=c]$ for all $c\in C^*$. On the other hand, suppose $\Phi^*\vdash_3\varphi[x:=c]$ for all $c\in C^*$, where $x\in fvar(\varphi)$. Let $n$ be minimal with the property $\varphi\in Fm(C_n)$. Then $\varphi[x:=c_{\varphi,x}]\in Fm(C_{n+1})$ and $c_{\varphi,x}\in C_{n+1}\smallsetminus C_n$. By construction, $\neg (\varphi^x)\in Y(C_n)\subseteq\Phi_{n+1}\subseteq\Phi^*$. Thus, $\Phi^*\vdash_3\neg (\varphi^x)$. Towards a contradiction suppose $\Phi^*\nvdash_3\forall x\varphi$. Since $\Phi^*$ is maximally consistent, $\Phi^*\vdash_3\neg\forall x\varphi$. Since $\Phi^*\vdash_3\varphi[x:=c]$ for all $c\in C^*$, we have in particular $\Phi^*\vdash_3\varphi[x:=c_{\varphi,x}]$. Thus, $\Phi^*\vdash_3\neg \forall x\varphi\wedge\varphi[x:=c_{\varphi,x}]$. Equivalently, $\Phi^*\vdash_3\neg(\neg\forall x\varphi\rightarrow\neg\varphi[x:=c_{\varphi,x}])$. That is, $\Phi^*\vdash_3\varphi^x$. This is a contradiction to $\Phi^*\vdash_3\neg (\varphi^x)$ and the consistency of $\Phi^*$. Therefore, $\Phi^*\vdash_3\forall x\varphi$. We have shown that $\Phi^*$ has the properties of a Henkin set. Let $(\mathcal{M}^*,\beta)$ be a normal model of the Henkin set $\Phi^*\subseteq Fm(C^*)$ and let $\varGamma^*\colon  C^*\rightarrow M$ be its Gamma-function. If we consider the restriction $\varGamma\colon C\rightarrow M$ of $\varGamma^*$ to $C\subseteq C^*$, then we get a normal model $\mathcal{M}$ w.r.t. the sublanguage $Fm(C)$, the reduct of $\mathcal{M}^*$. Obviously, $(\mathcal{M},\beta)\vDash\Phi$. Q.E.D.\\

If $\Phi\nvdash_3\varphi$, then using the Deduction Theorem (Lemma \ref{100}) one shows that $\Phi\cup\{\neg\varphi\}$ is consistent. The existence of a normal model of that set implies $\Phi\nVdash_3\varphi$. The Completeness Theorem follows.

\begin{theorem}[Completeness]\label{660}
For all $\Phi\cup\{\varphi\}\subseteq Fm(C)$: $\Phi\Vdash_3\varphi\Leftrightarrow\Phi\vdash_3\varphi$.
\end{theorem}

The result extends straightforwardly to Completeness Theorems for the systems $S4_\equiv^\forall$ and $S5_\equiv^\forall$ w.r.t. the above defined semantics.

\section{Propositional identity, strict equivalence and the Collapse Theorem}

Recall that by the Collapse Axiom we mean the scheme $(\square\varphi\wedge\square\psi)\rightarrow(\varphi\equiv\psi)$. This logical property can be expressed in algebraic terms in the following way: ``In every normal model, the smallest filter is $\{f_\top\}$."

\begin{lemma}\label{700}
Propositional identity w.r.t. a given interpretation $(\mathcal{M},\gamma)$ is a congruence relation containing alpha-congruence on $Fm(C)$.\footnote{By a congruence relation on $Fm(C)$ we mean an equivalence relation which is compatible with the connectives $\neg,\vee,\wedge,\rightarrow,\square,\equiv$ (but not necessarily with the quantifier $\forall$).} Strict equivalence w.r.t. a given interpretation is an equivalence relation on $Fm(C)$. Moreover, propositional identity refines strict equivalence. That is, 
\begin{equation*}
\Vdash_3(\varphi\equiv\psi)\rightarrow (\square(\varphi\rightarrow\psi)\wedge\square(\psi\rightarrow\varphi)).
\end{equation*}
\end{lemma}

\paragraph*{Proof.}
Given a model $(\mathcal{M},\gamma)$, it follows easily from model-theoretic properties that $\varphi\approx_i\psi :\Leftrightarrow (\mathcal{M},\gamma)\vDash\varphi\equiv\psi$ defines a congruence relation on $Fm(C)$ which contains alpha-congruence. Similarly, the relation $\varphi\approx_s\psi:\Leftrightarrow(\mathcal{M},\gamma)\vDash\square(\varphi\rightarrow\psi)\wedge\square(\psi\rightarrow\varphi)$ defines an equivalence relation. Now suppose $(\mathcal{M},\gamma)\vDash\varphi\equiv\psi$. This implies $\gamma(\varphi)=\gamma(\psi)$. Since $\square(\varphi\rightarrow\varphi)$ is valid, $f_\square(f_\rightarrow(\gamma(\varphi),\gamma(\varphi)))=f_\square(f_\rightarrow(\gamma(\varphi),\gamma(\psi)))\in TRUE$. That is, $(\mathcal{M},\gamma)\vDash \square(\varphi\rightarrow\psi)$. Similarly, one shows $(\mathcal{M},\gamma)\vDash \square(\psi\rightarrow\varphi)$. This shows the last assertion of the lemma. Q.E.D.\\

Note that strict equivalence is in general not a \textit{congruence} on $Fm(C)$. The reason for this fact is the identity connective: see the remarks after Definition \ref{200}.

If the relations of strict equivalence and propositional identity coincide, then the algebraic structure of models simplifies dramatically:

\begin{theorem}[Collapse Theorem]\label{720}
Let $\mathcal{M}$ be a normal model and $\le_\mathcal{M}$ its preorder. The following are equivalent:
\begin{enumerate}
\item $\mathcal{M}$ is a Boolean algebra and satisfies the Collapse Axiom. 
\item $\mathcal{M}$ is a Boolean algebra with $\mathit{NEC}=\{f_\top\}$.
\item $\le_\mathcal{M}$ is a partial order.
\item Strict equivalence coincides with propositional identity, that is:\\ $\mathcal{M}\vDash\forall x\forall y((x\equiv y)\leftrightarrow (\square(x\rightarrow y)\wedge\square(y\rightarrow x)))$. 
\end{enumerate}
\end{theorem}

\paragraph*{Proof.}
(i) $\Rightarrow$ (ii) is clear.\\
(ii) $\Rightarrow$ (iii): Let $\le$ be the lattice order. Then $f_\rightarrow(a,b)=f_\top\Leftrightarrow a\le b$, for all $a,b\in M$, as in any Boolean algebra. But under the condition $\mathit{NEC}=\{f_\top\}$, this is exactly the definition of the preorder $\le_\mathcal{M}$ in Definition \ref{200}.\\
(iii) $\Rightarrow$ (iv): If $(\mathcal{M},\gamma)\vDash \square(x\rightarrow y)\wedge\square(y\rightarrow x)$, then $\gamma(x)\le_\mathcal{M}\gamma(y)$ and $\gamma(y)\le_\mathcal{M}\gamma(x)$. Since $\le_\mathcal{M}$ is a partial order, $\gamma(x)=\gamma(y)$. Thus, $(\mathcal{M},\gamma)\vDash x\equiv y$.\\
(iv) $\Rightarrow$ (i): $\mathcal{M}$ is a Boolean prelattice with preorder $\le_\mathcal{M}$ given by $a\le_\mathcal{M} b\Leftrightarrow f_\rightarrow(a,b)\in \mathit{NEC}$. Suppose $a\approx_\mathcal{M}b$, i.e. $a\le_\mathcal{M}b$ and $b\le_\mathcal{M} a$. If we assign $a,b$ to the variables $x,y$, respectively, then condition (iv) yields $a=b$. That is, $\approx_\mathcal{M}$ is the identity on $M$ and the quotient algebra of $\mathcal{M}$ modulo $\approx_\mathcal{M}$ is $\mathcal{M}$ itself, which, by Definition \ref{140}, must be a Boolean algebra. Moreover, by Lemma \ref{170}, $\mathit{NEC}=\{a\in M\mid a\approx_\mathcal{M} f_\top\}$. Since $\approx_\mathcal{M}$ is the identity, the Collapse Axiom follows. Q.E.D.\\

Note that if the normal model $\mathcal{M}$ is a Boolean algebra, then its lattice order $\le$ is not necessarily the preorder $\le_\mathcal{M}$. In other words, the set $\mathit{NEC}$, which is a filter w.r.t. $\le_\mathcal{M}$, may strictly extend the (smallest) lattice filter $\{f_\top\}$ of the Boolean algebra. Since the lattice order $\le$ refines $\le_\mathcal{M}$, $\mathit{NEC}$ is also a lattice filter. The lattice order coincides with $\le_\mathcal{M}$ if and only if the Boolean algebra $\mathcal{M}$ satisfies the Collapse Axiom. Similarly, the condition of a model $\mathcal{M}$ to satisfy the Collapse Axiom is not sufficient for $\mathcal{M}$ being a Boolean algebra: $\le_\mathcal{M}$ may be not anti-symmetric.\\ 

The models of the modal $\mathit{SCI}$-theories studied in \cite{sus2} satisfy the properties (i)--(iv) of the Collapse Theorem. Also the models of the non-Fregean logic developed by Ishi \cite{ish1, ish2} are Boolean algebras and satisfy the Collapse Axiom (the identity connective of that logic, however, satisfies in general not all $\mathit{SCI}$-axioms of propositional identity).

\begin{theorem}\label{740}
We consider here the language $Fm_m$ of basic modal propositional logic. If we introduce an identity connective defining
\begin{equation*}
\varphi\equiv\psi:=\square(\varphi\rightarrow\psi)\wedge\square(\psi\rightarrow\varphi),
\end{equation*}
then the axiom schemes of propositional identity (v)--(vii) of $\mathbb{AX}$ are derivable in $S3$.\footnote{In this quantifier-free context, we may replace axiom (v) by the stronger (v'): $\varphi\equiv\varphi$.} That is, propositional identity is definable by strict equivalence in $S3$.
\end{theorem}

\paragraph*{Proof.}
Suppose a connective $\equiv$ is defined in that way. We consider derivations in modal logic $S3$. Since $\square(\varphi\rightarrow\varphi)$ is derivable (by Axiom Necessitation), we get $\varphi\equiv\varphi$, i.e. axiom (v') of propositional identity. Axiom (vi) derives from axiom (ii). In order to prove that axiom (vii) is derivable it suffices to show that the following are theorems:
\begin{itemize}
\item $(\varphi\equiv\psi)\rightarrow (\neg\varphi\equiv\neg\psi)$
\item $((\varphi\equiv\psi)\wedge (\varphi'\equiv\psi'))\rightarrow ((\varphi\rightarrow\varphi')\equiv (\psi\rightarrow\psi'))$
\item $((\varphi\equiv\psi)\wedge (\varphi'\equiv\psi'))\rightarrow ((\varphi\wedge\varphi')\equiv (\psi\wedge\psi'))$
\item $((\varphi\equiv\psi)\wedge (\varphi'\equiv\psi'))\rightarrow ((\varphi\vee\varphi')\equiv (\psi\vee\psi'))$
\item $((\varphi\equiv\psi)\wedge (\varphi'\equiv\psi'))\rightarrow ((\varphi\equiv\varphi')\equiv (\psi\equiv\psi'))$
\item $(\varphi\equiv\psi)\rightarrow (\square\varphi\equiv\square\psi)$
\end{itemize}

It is known that $(\square\varphi\wedge\square\psi)\leftrightarrow\square(\varphi\wedge\psi)$ is a theorem of $S3$. Hence, strict equivalence between $\varphi$ and $\psi$ can be expressed by $\square(\varphi\leftrightarrow\psi)$. 
By propositional logic and rule AN we get $\square((\varphi\leftrightarrow\psi)\rightarrow(\neg\varphi\leftrightarrow\neg\psi))$. Axiom (iii) and MP then yield the first theorem above. Similarly, we get the second, third and fourth theorem. Let us look at formula number 5. By propositional logic and AN: $\square(((\varphi\leftrightarrow\psi)\wedge(\varphi'\leftrightarrow\psi'))\rightarrow((\varphi\leftrightarrow\varphi')\leftrightarrow(\psi\leftrightarrow\psi')))$. By axiom (iii) and MP:
$\square((\varphi\leftrightarrow\psi)\wedge(\varphi'\leftrightarrow\psi'))\rightarrow\square((\varphi\leftrightarrow\varphi')\leftrightarrow(\psi\leftrightarrow\psi'))$. By axiom (iv) and transitivity of implication:
$\square((\varphi\leftrightarrow\psi)\wedge(\varphi'\leftrightarrow\psi'))\rightarrow\square(\square(\varphi\leftrightarrow\varphi')\leftrightarrow\square(\psi\leftrightarrow\psi'))$. This yields the fifth theorem. Finally, by axiom (iv) we have $\square(\varphi\rightarrow\psi)\rightarrow\square(\square\varphi\rightarrow\square\psi)$ and $\square(\psi\rightarrow\varphi)\rightarrow\square(\square\psi\rightarrow\square\varphi)$. Hence, $(\square(\varphi\rightarrow\psi)\wedge\square(\psi\rightarrow\varphi)) \rightarrow (\square(\square\varphi\rightarrow\square\psi)\wedge\square(\square\psi\rightarrow\square\varphi))$. From this one easily derives the last theorem. The scheme of axiom (vii) now follows by induction on formulas. Q.E.D.

\begin{corollary}
$S3$ is the weakest Lewis modal system in which propositional identity is definable by strict equivalence.
\end{corollary}

\paragraph*{Proof.}
We saw that in $S3$ all axioms of propositional identity can be derived if one defines propositional identity by strict equivalence. A particular axiom of propositional identity is the following: $(\varphi\equiv\psi)\rightarrow(\square\varphi\equiv\square\psi)$, i.e., $(\square(\varphi\rightarrow\psi)\wedge\square(\psi\rightarrow\varphi))\rightarrow (\square(\square\varphi\rightarrow\square\psi)\wedge\square(\square\psi\rightarrow\square\varphi))$. This, however, is not a theorem of the weaker Lewis system $S2$ as one can show by constructing a Kripke model of $S2$ (i.e., a Kripke model with at least one normal world and reflexive accessibility relation) where that formula is not true. Q.E.D.

\section{Representation theorems}

K. Fine \cite{fin} extends normal modal logics by axioms for propositional quantifiers and studies several conditions which can be imposed upon the set of propositions. A natural condition, trivially satisfied in our denotational approach, is that propositions ``are closed under formulas", i.e., each formula under any valuation denotes (``interprets") a proposition. In particular, propositions are closed under Boolean operations. We define here a $S3\pi$-frame as a triple $\mathcal{F}=(W,N,R,P)$, where $W$ is a set of worlds, $N\subseteq W$ is a non-empty set of normal worlds, $R\subseteq W\times W$ is a reflexive and transitive accessibility relation, and $P\subseteq Pow(W)$ is the set of propositions (``closed under formulas"). In particular, $\varnothing, W\in P$. We may assume here that the only world accessible from a non-normal world $w$ is $w$ itself. This will be helpful for the definition of \textit{proposition} in the context of non-normal modal logic S3. We work with the language $Fm(C_0)$ where $C_0=\{\bot,\top\}$. A valuation is a function $g\colon V\rightarrow P$ which extends to the set of constants such that $g(\bot):=\varnothing$ and $g(\top):= W$. If $g,g'$ are valuations such that $g(y)=g'(y)$ for all $y\in V\smallsetminus\{x\}$, then we write $g=_x g'$. The satisfaction relation for a normal world $w\in N$ is defined as follows:
\begin{equation*}
\begin{split}
&(w,g)\vDash x :\Leftrightarrow w\in g(x),\text{ for }x\in V\\
&(w,g)\vDash c :\Leftrightarrow w\in g(c),\text{ for }c\in C_0\\
&(w,g)\vDash\varphi\vee\psi :\Leftrightarrow (w,g)\vDash\varphi\text{ or }(w,g)\vDash\psi\\
&(w,g)\vDash\varphi\wedge\psi :\Leftrightarrow (w,g)\vDash\varphi\text{ and }(w,g)\vDash\psi\\
&(w,g)\vDash\varphi\rightarrow\psi :\Leftrightarrow (w,g)\nvDash\varphi\text{ or }(w,g)\vDash\psi\\
&(w,g)\vDash\neg\varphi :\Leftrightarrow (w,g)\nvDash\varphi\\
&(w,g)\vDash\square\varphi :\Leftrightarrow (w',g)\vDash\varphi,\text{ for all }w'\text{ such that }wRw'\\
&(w,g)\vDash\varphi\equiv\psi :\Leftrightarrow (w',g)\vDash\varphi\text{ iff }(w',g)\vDash\psi,\text{ for all }w'\text{ such that }wRw'\\
&(w,g)\vDash\forall x\varphi :\Leftrightarrow (w,g')\vDash\varphi\text{ for all valuations }g'\text{ such that }g'=_x g 
\end{split}
\end{equation*}

The satisfaction relation for a non-normal world $w\in W\smallsetminus N$ is given in the same way except for the condition concerning the modal operator which is replaced by the following:
\begin{equation*}
(w,g)\nvDash\square\varphi
\end{equation*}

Let $S3\pi$ be the set of formulas true at all \textit{normal} worlds in all $S3\pi$-frames under all valuations. If we consider those frames where $N=W$, then we obtain the theory $S4\pi$. $S5\pi$ results from $S4\pi$ by imposing the additional condition that $R$ in each frame is an equivalence relation. This is essentially the same way as the theories $S4\pi$ and $S5\pi$ are defined in \cite{fin}. Of course, our theories contain, in addition, theorems with identity connective (this connective is not an element of the language considered by Fine \cite{fin}). Note that all axioms of $\mathbb{AX}$ belong to $S3\pi$. One also easily checks that \\
$(\varphi\equiv\psi)\leftrightarrow\square(\varphi\leftrightarrow\psi)$\\
$(\varphi\equiv\psi)\rightarrow\square(\varphi\equiv\psi)$ \\
belong to $S3\pi$. Recall that the latter is also derivable from $\mathbb{AX}$ (see Lemma \ref{130}). The former, however, is valid iff the Collapse Axiom holds (see Theorem \ref{720}). Note that $\square(\varphi\leftrightarrow\psi)\rightarrow\square\square(\varphi\leftrightarrow\psi)$ is not a theorem of $S3\pi$. So we cannot replace $\varphi\equiv\psi$ by $\square(\varphi\leftrightarrow\psi)$ in every context (both formulas are equivalent in normal worlds but they do not necessarily denote the same proposition).  \\

In standard modal logic, a proposition is usually regarded as a set of possible worlds. Relative to a given world $w$ of a given frame one may regard the proposition denoted by $\varphi$ as the set of those worlds which are accessible from $w$ and where $\varphi$ is true. Accordingly, two formulas $\varphi$ and $\psi$ denote the same proposition at world $w$ iff $\varphi\equiv\psi$ is true at $w$.

\begin{theorem}\label{800}
Let $k\in\{3,4,5\}$, let $\mathcal{F}=(W,N,R,P)$ be a $Sk\pi$-frame and $C_0=\{\bot,\top\}$. For every world $w\in W$ and every valuation $g\colon V\rightarrow P$ there exist a $Sk_\equiv^\forall$-model $\mathcal{M}$ satisfying the Collapse Axiom and an assignment $\gamma\colon V\rightarrow M$ such that for all $\varphi,\psi\in Fm(C_0)$ the following holds:
\begin{equation*}
(\mathcal{M},\gamma)\vDash\varphi\Leftrightarrow (w,g)\vDash\varphi.
\end{equation*}
In particular, $(\mathcal{M},\gamma)\vDash\varphi\equiv\psi\Leftrightarrow (w,g)\vDash\varphi\equiv\psi$. That is, $\varphi$ and $\psi$ denote the same proposition in $\mathcal{M}$ iff they denote the same proposition in $\mathcal{F}$ at world $w$. Thus, the concept of a \textit{proposition} as the denotation of a formula in model $\mathcal{M}$ and the modal concept of a \textit{proposition} as a set of possible worlds are equivalent.
\end{theorem}

\paragraph*{Proof.}
For each $p\in P$ let $c_p$ be a constant symbol such that $p\neq q$ implies $c_p\neq c_q$. Put $C:=\{c_p\mid p\in P\}$. We may assume that $\bot,\top\in C$. A valuation $g\colon V\rightarrow P$ now extends to a function on $V\cup C$ such that $g(c_p)=p$. The second clause of the truth definition above says: $(w,g)\vDash c :\Leftrightarrow w\in g(c)$, where $c$ is now any element of $C$. By induction on formulas, simultaneously for all valuations, one shows the following facts:\\
\textit{Coincidence Lemma}: For all $w\in W$ and all $\varphi\in Fm(C)$, if $g(x)=g'(x)$ for all $x\in fvar(\varphi)$, then $(w,g)\vDash\varphi\Leftrightarrow (w,g')\vDash\varphi$.\\
\textit{Substitution Lemma}: For any $w\in W$, $p_1,...,p_n \in P$, $x_1,...,x_n\in V$, $\varphi\in Fm(C)$ and any valuation $g$, $(w,g_{x_1,...x_n}^{p_1,...,p_n})\vDash\varphi\Leftrightarrow (w,g)\vDash\varphi[x_1:=c_{p_1},...,x_2:=c_{p_n}]$.\\
As a consequence we obtain the following:

\begin{equation}\label{500}
(w,g)\vDash \forall x\varphi\Leftrightarrow (w,g_x^p)\vDash\varphi\text{ for all }p\in P\Leftrightarrow (w,g)\vDash\varphi[x:=c_p]\text{ for all }p\in P.
\end{equation}
Now let $w\in W$ and let $g\colon V\rightarrow P$ be a valuation. Define the relation $\approx$ on $Fm(C)$ by $\varphi\approx\psi:\Leftrightarrow(w,g)\vDash\varphi\equiv\psi$. One easily checks that $\approx$ is a congruence relation on $Fm(C)$. For $\varphi\in Fm(C)$ let $\overline{\varphi}$ be the equivalence class of $\varphi$ modulo $\approx$. Every formula denotes a proposition (in the terminology of \cite{fin}, ``$P$ is closed under formulas"). Thus, for each $\varphi$ there is a constant $c\in C$ such that $\varphi\approx c$. In fact, we may choose $c=c_p$ if $\varphi$ denotes the proposition $p\in P$ under the valuation $g$. Define 
\begin{equation*}
\begin{split}
&M:=\{\overline{\varphi}\mid\varphi\in Fm(C)\}=\{\overline{c}\mid c\in C\}\\
&\mathit{TRUE}:=\{\overline{\varphi}\mid (w,g)\vDash\varphi\}\\
&\mathit{NEC}:=\{\overline{\varphi}\mid  (w,g)\vDash\square\varphi\}\\
&f_\neg(\overline{\varphi}):=\overline{\neg\varphi}, f_\square(\overline{\varphi})=\overline{\square\varphi},f_\top:=\overline{\top}, f_\bot:=\overline{\bot},\text{ and }f_@(\overline{\varphi},\overline{\psi}):=\overline{\varphi @\psi}
\end{split}
\end{equation*}
for $@\in\{\vee, \wedge, \rightarrow,\equiv\}$. The Collapse Axiom holds and $\mathit{NEC}=\{f_\top\}$. The above sets and operations are well-defined and form a Boolean algebra $\mathcal{M}$. We define the Gamma-function by $\varGamma(c):=\overline{c}$. Finally, the higher-order function $f_\forall\colon M^M\rightarrow M$ is given by 
\begin{equation*}
\begin{split}
f_\forall(t)=:
\begin{cases}
&\overline{\forall x\varphi},\text{ if there is a }\varphi\text{ such that }t(\overline{c})=\overline{\varphi[x:=c]}\text{ for all }c\in C\\
&f_\top,\text{ if such a formula }\varphi\text{ does not exist}
\end{cases}
\end{split}
\end{equation*}

Now we may argue in a similar way as in the proof of Theorem \ref{620}, where a model for a Henkin set is constructed. By \eqref{500}, $\Phi=\{\varphi\in Fm(C)\mid (w,g)\vDash\varphi\}$ has in fact the properties of a Henkin set. We show that $f_\forall$ is well-defined. Suppose $t\in M^M$ such that $\overline{\varphi[x:=c]}=t(\overline{c})=\overline{\psi[y:=c]}$ for two formulas $\varphi,\psi\in Fm(C)$ and for all $c\in C$. Without lost of generality, we may assume that $x\notin var(\psi)$. Then $\varphi[x:=c]\approx \psi[y:=c]= (\psi[y:=x])[x:=c]$ for all $c\in C$. That is, $(w,g)\vDash\square((\varphi\leftrightarrow\psi[y:=x])[x:=c])$ for all $c\in C$. By \eqref{500}, $(w,g)\vDash\forall x \square((\varphi\leftrightarrow\psi[y:=x])$. The Kripke semantics implies: $(w,g)\vDash\square(\forall x\varphi\leftrightarrow\forall x\psi[y:=x])$. That is, $\overline{\forall x\varphi}=\overline{\forall x\psi[y:=x]}$. Note that $\forall x\psi[y:=x]$ and $\forall y\psi$ are alpha-congruent. Thus, $f_\forall(t)=\overline{\forall x\varphi}=\overline{\forall y\psi}$ and $f_\forall$ is well-defined. One verifies that all conditions of a $S3^\forall_\equiv$-model are satisfied. For instance, condition (iv)(k) holds because the Barcan formula and its converse belong to $S3\pi$. Let $\gamma\colon V\rightarrow M$ be the assignment defined by $x\rightarrow\overline{x}$. In the same way as in Claim 2 of Theorem \ref{620} one shows by induction of the quantifier-rank that $\gamma(\varphi)=\overline{\varphi}$ for all $\varphi\in Fm(C)$. Then for every $\varphi\in Fm(C)$:
\begin{equation*}
(\mathcal{M},\gamma)\vDash\varphi\Leftrightarrow\gamma(\varphi)=\overline{\varphi}\in TRUE\Leftrightarrow (w,g)\vDash\varphi.
\end{equation*}
Finally, we consider the ``reducts" of both models (i.e., the restrictions of the Gamma-function, of the valuation $g$, respectively) to the sublanguage $Fm(C_0)\subseteq Fm(C)$. This yields the assertions. Note that $\mathcal{M}$ is the two-element Boolean algebra if $w$ is a non-normal world. Q.E.D.

\begin{lemma}\label{810}
Let $F$ be a filter of a $S3_\equiv^\forall$-model $\mathcal{M}$. Then $F$ is the intersection of all ultrafilters that extend $F$.
\end{lemma}

\paragraph*{Proof.}
Let $X=\bigcap\{U\subseteq M\mid U\supseteq F$ is an ultrafilter$\}$. Then $F\subseteq X$. Suppose there is $a\in X\smallsetminus F$. Using Zorn's Lemma (or an appropriate weaker principle) one shows that $F$ extends to a maximal filter (i.e., an ultrafilter) which does not contain $a$. We get $a\notin X$, a contradiction. Hence, $F=X$, i.e., $F$ is the meet of all ultrafilters extending $F$. Q.E.D.\\

Some parts of the next result have parallels to the J\'onsson-Tarski Theorem which essentially says that a Boolean algebra with operators is embeddable in the full complex algebra of its ultrafilter frame (see, e.g., \cite{bla} for a detailed discussion). In the proof of the following Theorem \ref{820} we shall construct a desired Kripke model from the ultrafilters of a given $Sm_\equiv^\forall$-model, where $m\in\{4,5\}$, such that the same formulas are satisfied. We were unable to prove the theorem for arbitrary $S3_\equiv^\forall$-models. Note that also the J\'onsson-Tarski Theorem is applicable only to \textit{normal} modal logics.

Recall that by $Fm_m$ we denote the set of formulas of pure modal logic (without identity connective and without quantifier).

\begin{theorem}\label{820}
Let $\mathcal{M}$ be a $S4_{\equiv}^\forall$-model and let $\gamma\colon V\rightarrow M$ be an assignment. There exist a frame $(W,R)$ of modal logic $S4$, a valuation $g\colon V\rightarrow Pow(W)$ and a world $w\in W$ such that for all $\varphi,\psi\in Fm_m$: 
\begin{equation}\label{4500}
\begin{split}
&(\mathcal{M},\gamma)\vDash\varphi\Leftrightarrow (w,g)\vDash\varphi, \text{ and}\\
&(\mathcal{M},\gamma)\vDash\varphi\equiv\psi\Rightarrow (w,g)\vDash\square(\varphi\leftrightarrow\psi).
\end{split}
\end{equation}
Moreover, if the model $\mathcal{M}$ satisfies the Collapse Axiom and is a Boolean algebra, then the implication in the second line of \eqref{4500} can be replaced by a biconditional $\Leftrightarrow$, i.e., $\varphi,\psi\in Fm_m$ denote the same proposition in $\mathcal{M}$ under $\gamma$ iff they denote the same proposition at world $w$ under valuation $g$.
\end{theorem}

\paragraph*{Proof.}
Let $\mathit{TRUE}$, $\mathit{NEC}$ be the sets of true, necessary propositions, respectively, $\le_\mathcal{M}$ the induced preorder of $\mathcal{M}$ and $W:=\{T\mid T$ is an ultrafilter w.r.t. $\le_\mathcal{M}\}$. Then $\mathit{TRUE}\in W$. For each $a\in M$ let $|a|:=\{T\mid a\in T\in W\}$. Define $P=\{|a|\mid a\in M\}$, the set of propositions for the desired Kripke model. For $T\in W$ let $\mathit{NEC_T}:=\{a\in M\mid f_\square(a)\in T\}$ and define the relation $\le_T$ by $a\le_T b :\Leftrightarrow f_\rightarrow (a,b)\in \mathit{NEC_T}$.\\
\textbf{Claim 1}: For each $T\in W$, $\mathit{NEC}\subseteq T$.\\
\textit{Proof of the claim.} By Lemma \ref{170}, $\mathit{NEC}=\{a\in M\mid a\approx_\mathcal{M} f_\top\}$ and $\mathit{NEC}$ is the smallest filter.\\
\textbf{Claim 2}: For each $T\in W$, $\mathit{NEC}_T\subseteq T$. In particular, $T$ is an ultrafilter w.r.t. $\le_T$.\\
\textit{Proof of the claim.} Let $a\in \mathit{NEC}_T$. By definition, $f_\square(a)\in T$. Since $f_\square(a)\le_\mathcal{M} a$ and $T$ is a filter, we get $a\in T$. This shows the first part of the claim. We have $f_\rightarrow(a,b)\approx_\mathcal{M} f_\vee(f_\neg(a),b))$ because $\mathcal{M}$ is a Boolean prealgebra. Then $a\in T$ and $a\le_T b$ imply $b\in T$.\\
\textbf{Claim 3}: For each $T\in W$, $\mathit{NEC}\subseteq \mathit{NEC}_T$. In particular, $\le_\mathcal{M}$ refines $\le_T$.\\
\textit{Proof of the claim.} Let $a\in \mathit{NEC}$. That is, $f_\square(a)\in \mathit{TRUE}$. $f_\square(a)\le_\mathcal{M} f_\square (f_\square(a))$ because $\mathcal{M}$ is a $S4_\equiv^\forall$-model. Since $\mathit{TRUE}$ is a filter, we get $f_\square(a)\in \mathit{NEC}\subseteq T$. By definition, $a\in \mathit{NEC_T}$.\\
\textbf{Claim 4}: Every ultrafilter with respect to $\le_T$ belongs to $W$.\\
\textit{Proof of the claim.} By Claim 3, $\le_\mathcal{M}$ refines $\le_T$.\\
\textbf{Claim 5}: For each $T\in W$, if $a\le_T b$ and $a\in \mathit{NEC_T}$, then $b\in \mathit{NEC_T}$.\\
\textit{Proof of the claim.} Let $a\le_T b$ and $a\in \mathit{NEC_T}$. Then $f_\square(f_\rightarrow(a,b))\in T$ and $f_\square(a)\in T$. $f_\square(f_\rightarrow(a,b))\le_\mathcal{M} f_\rightarrow(f_\square(a),f_\square(b))$ and $T$ is an ultrafilter. Thus, $f_\rightarrow(f_\square(a),f_\square(b))\in T$ and finally $f_\square(b)\in T$. That is, $b\in\mathit{NEC_T}$.\\
\textbf{Claim 6}: For each $T\in W$, if $a,b\in \mathit{NEC_T}$, then $f_\wedge(a,b)\in \mathit{NEC_T}$.\\
\textit{Proof of the claim.} Let $a,b\in \mathit{NEC_T}$. Then $f_\square(a), f_\square(b)\in T$ and therefore $f_\wedge(f_\square(a)$, $f_\square(b))\in T$. Note that $\varphi:=x\rightarrow (y\rightarrow (x\wedge y))$ is a propositional tautology. By Axiom Necessitation, $\square\varphi$ is a theorem. By soundness, $\square\varphi$ is valid. Choose an assignment $x\mapsto a$, $y\mapsto b$. This shows $a\le_\mathcal{M} f_\rightarrow(b,f_\wedge(a,b))$. Since $a\in \mathit{NEC_T}$, Claim 3 and Claim 5 yield $f_\rightarrow(b,f_\wedge(a,b))\in \mathit{NEC_T}$. That is, $b\le_T f_\wedge(a,b)$. By Claim 5, $f_\wedge(a,b)\in \mathit{NEC_T}$.\\ 
\textbf{Claim 7}: For each $T\in W$, $\mathit{NEC_T}$ is the smallest filter w.r.t. $\le_T$.\\
\textit{Proof of the claim.} By Claim 5 and Claim 6, $\mathit{NEC_T}$ is a filter w.r.t. $\le_T$. Similarly as in Lemma \ref{170} one shows that $\mathit{NEC_T}=\{a\mid a\approx_T f_\top\}$, where $a\approx_T b:\Leftrightarrow (a\le_T b$ and $b\le_T a)$. Any filter contains $f_\top$ and the claim follows.\\
\textbf{Claim 8}: For each $T\in W$, $\mathit{NEC_T}=\bigcap\{T'\in W\mid \mathit{NEC_T}\subseteq T'\}$.\\
\textit{Proof of the claim.} Since $\mathit{NEC_T}$ is the smallest filter w.r.t. $\le_T$, it is the intersection of all ultrafilters w.r.t. $\le_T$. By Claim 4, all those ultrafilters belong to $W$ and the claim follows.\\

We define the accessibility relation $R$ on $W$ by:
\begin{equation*}
TRT':\Leftrightarrow \mathit{NEC_T}\subseteq T'.
\end{equation*}
It is clear that $R$ is reflexive. Suppose $TRT'RT''$. Let $a\in\mathit{NEC_T}$. Since we are dealing with a $S4_\equiv^\forall$-model, $f_\square(a)\in \mathit{NEC_T}\subseteq T'$. Then $a\in \mathit{NEC_{T'}}\subseteq T''$. Hence, $\mathit{NEC_T}\subseteq T''$. This shows that $R$ is transitive. Note that each $\mathit{NEC_T}$ is non-empty because $\mathit{NEC}\subseteq \mathit{NEC_T}$. Hence, there are no non-normal worlds in $W$. Thus, $(W,R)$ is a frame of modal logic $S4$. For a given assignment $\beta\colon V\rightarrow M$ of model $\mathcal{M}$ we define the valuation $g_\beta\colon V\rightarrow P$ by $g_\beta(x):=|\beta(x)|$.\\
\textbf{Claim 9}: For any $\varphi\in Fm_m$, any assignment $\beta\colon V\rightarrow M$ of model $\mathcal{M}$ and any world $T\in W$:
\begin{equation*}
(T,g_\beta)\vDash\varphi\Leftrightarrow\beta(\varphi)\in T.
\end{equation*}
The claim follows by induction on $\varphi\in Fm_m$. The basis case $\varphi=x$ is true by the definition of $g_\beta$: $(T,g_\beta)\vDash x\Leftrightarrow T\in g_\beta(x)=|\beta(x)|\Leftrightarrow\beta(x)\in T$. Most of the remaining cases now follow straightforwardly from the induction hypothesis and the definition of an assignment. We show the case $\varphi=\square\psi$:
\begin{equation*}
\begin{split}
(T,g_\beta)\vDash\square\psi &\Leftrightarrow(T',g_\beta)\vDash\psi,\text{ for all }T'\in W\text{ with }TRT'\\
&\Leftrightarrow \beta(\psi)\in T',\text{ for all }T'\in W\text{ with }TRT',\text{ by induction hypothesis}\\
&\Leftrightarrow \beta(\psi)\in\bigcap\{T'\in W\mid NEC_T\subseteq T'\},\text{ by definition of }R\\
&\Leftrightarrow\beta(\psi)\in \mathit{NEC_T},\text{ by Claim 8 }\\
&\Leftrightarrow f_\square(\beta(\psi))\in T,\text{ by definition of }\mathit{NEC_T}\\
&\Leftrightarrow\beta(\square\psi)\in T,\text{ by definition of an assignment}
\end{split}
\end{equation*}
Thus, Claim 9 is true. We consider the world $\mathit{TRUE}\in W$, the given assignment $\gamma:V\rightarrow M$ and the valuation $g_\gamma$.\footnote{Note that $\mathit{NEC}=\mathit{NEC}_\mathit{TRUE}$. Thus, by Claim 1, the world $\mathit{TRUE}$ accesses every $T\in W$.} Then for all $\varphi\in Fm_m$:
\begin{equation*}\label{5000}
(\mathit{TRUE},g_\gamma)\vDash\varphi\Leftrightarrow\gamma(\varphi)\in \mathit{TRUE}\Leftrightarrow (\mathcal{M},\gamma)\vDash\varphi.
\end{equation*}
This shows the first part of \eqref{4500}. Now suppose $(\mathcal{M},\gamma)\vDash\varphi\equiv\psi$ for $\varphi,\psi\in Fm_m$. Then $\gamma(\varphi)=\gamma(\psi)$. Thus, $\gamma(\varphi)\in T$ iff $\gamma(\psi)\in T$, for each $T\in W$. Then from Claim 9 it follows that $(\mathit{TRUE},g_\gamma)\vDash\square(\varphi\leftrightarrow\psi)$.

Finally, suppose $\mathcal{M}$ is a Boolean algebra that satisfies the Collapse Axiom. Then, by Theorem \ref{720}, propositional identity $\varphi\equiv\psi$ is given by strict equivalence $\square(\varphi\leftrightarrow\psi)$. The last assertion of the theorem now follows from the first line of \eqref{4500}. Q.E.D.

\begin{corollary}
If the model $\mathcal{M}$ in Theorem \ref{820} is a $S5_\equiv^\forall$-model, then we obtain a Kripke model $(W,R,g_\gamma)$  of modal logic $S5$ such that the assertions of the theorem remain true.
\end{corollary}

\paragraph*{Proof.}
The Claims 1--8 in the proof of Theorem \ref{820} remain true. Moreover, Claim 3 can be replaced by the stronger\\
\textbf{Claim 3'}: For each $T\in W$, $\mathit{NEC} = \mathit{NEC_T}$.\\
\textit{Proof of the Claim}. By Claim 3, $\mathit{NEC}\subseteq \mathit{NEC_T}$. Now suppose $a\notin \mathit{NEC}$. Then $f_\neg (f_\square (a))\in \mathit{TRUE}$. Since $\mathcal{M}$ is a $S5_\equiv^\forall$-model, $f_\square(f_\neg (f_\square a))\in \mathit{TRUE}$, that is, $f_\neg (f_\square(a))\in \mathit{NEC}\subseteq T$. Thus, $f_\square(a)\notin T$ and $a\notin \mathit{NEC_T}$. Hence, $\mathit{NEC_T}\subseteq \mathit{NEC}$ and therefore $\mathit{NEC}=\mathit{NEC_T}$.\\
The accessibility relation $R$ on $W$ is given as before. Then by Claim 3', $\mathit{NEC_T}=\mathit{NEC}=\mathit{NEC_{T'}}$ for any worlds $T, T'\in W$. Thus, all worlds of $W$ are related by $R$, and $R$ is an equivalence relation. Then $(W,R)$ is a frame of modal logic $S5$. Also Claim 9 is true. The assertion now follows in the same way as in the proof of the theorem. Q.E.D.

\begin{corollary}[Conservative extension]\label{830}
Our denotational semantics captures the standard modal systems $S3$--$S5$ in the following sense. For any $\varphi\in Fm_m$ and $k\in\{3,4,5\}$, $\varphi$ is a theorem of $Sk_\equiv^\forall$ iff $\varphi$ is a theorem of modal system $Sk$. Consequently, the theory $Sk_\equiv^\forall$ is a conservative extension of modal system $Sk$.
\end{corollary}

\paragraph*{Proof.}
$Sk_\equiv^\forall$ contains all axioms of $Sk$. If $k\in\{4,5\}$, then, by Lemma \ref{120}, also the Necessitation Rule is derivable. Thus, every theorem of $Sk$ is a theorem of $Sk_\equiv^\forall$, for $k=3,4,5$. Now suppose $\varphi\in Fm_m$ is not a theorem of $Sk$. Then there is a Kripke model of system $Sk$ with a \textit{normal} world $w$ and valuation $g\colon V\rightarrow Pow(W)$ such that $(w,g)\vDash\neg\varphi$. That Kripke model can be seen as a frame $(W,N,R,P)$  with $P=Pow(W)$. By Theorem \ref{800}, there is a normal $Sk_\equiv^\forall$-model $\mathcal{M}$ and an assignment $\gamma$ such that $(\mathcal{M},\gamma)\vDash\neg\varphi$. By soundness, $\varphi$ cannot be a theorem of $Sk_\equiv^\forall$. Q.E.D.

\section{A simpler and more intensional semantics}

$\mathbb{AX}$ contains the scheme (viii), $\forall x (\varphi\equiv\psi)\rightarrow (\forall x\varphi\equiv\forall x\psi)$, which represents an \textit{extensional} principle. It can be read as follows: ``Two definable functions are equal if they have the same extensions (the same graphs)". Our aim is to relax such extensional constraints whenever this is possible and meaningful. In fact, we are able to define a weaker semantics such that axiom scheme (viii) as well as the Barcan formula can be avoided. \\

Let $\mathbb{AX}^-$ be the set of axioms which is given by the smallest set that contains all formulas (i)--(vii) and (ix)--(xii) of $\mathbb{AX}$ and is closed under the following condition: If $\varphi\in\mathbb{AX}^-$ and $x\in fvar(\varphi)$, then $\forall x\varphi\in\mathbb{AX}^-$.

As before, an assignment of a model with universe $M$ is a function $\gamma\colon V\rightarrow M$. In contrast to the denotational semantics of the first kind, however, there is no canonical way to extend $\gamma$ to a function $\gamma\colon Fm(C)\rightarrow M$. In fact, there is no explicitly given algebraic structure on the universe of a model although parts of such structure can be restored. Instead of an explicit algebraic structure, there are certain structural conditions concerning assignments and substitutions. This style of semantics was designed in \cite{str} and has been further developed in \cite{zei} and \cite{lewigpl}. We shall adopt some technical machinery coming from the last two works, with some improvements and simplifications.

\begin{definition}
A simple model $\mathcal{M}=(M,\mathit{TRUE},\mathit{NEC},\varGamma)$ is given by a non-empty propositional universe $M$, sets $\mathit{NEC}\subseteq \mathit{TRUE}\subseteq M$ and a function $\varGamma\colon C\rightarrow M$ such that the following conditions are satisfied.\\
\textbf{Structural properties}:\footnote{In \cite{lewigpl}, the Gamma-function is a function $\varGamma\colon Fm(C)\times M^V\rightarrow M$ which extends any given assignment $\gamma\in M^V$ and maps any formula $\varphi$ to a proposition $\varGamma(\varphi,\gamma)\in M$. The present definition is equivalent to the definition given in \cite{lewigpl}. The connection is given by: ``$\gamma(\varphi)=\varGamma(\varphi,\gamma)$".}
\begin{itemize}
\item $\gamma(c)=\varGamma(c)$ for every assignment $\gamma\colon V\rightarrow M$ and every $c\in C$
\item If $\gamma,\gamma'\colon V\rightarrow M$ are assignments with $\gamma(x)=\gamma'(x)$ for all $x\in fvar(\varphi)$, then $\gamma(\varphi)=\gamma'(\varphi)$, for any $\varphi\in Fm(C)$. (Coincidence Property)
\item If $\sigma\colon V\rightarrow Fm(C)$ is a substitution, $\gamma\colon V\rightarrow M$ is an assignment, and $\gamma\sigma\colon V\rightarrow M$ is the assignment defined by $x\mapsto\gamma(\sigma(x))$, then $\gamma(\varphi[\sigma])=\gamma\sigma(\varphi)$, for any $\varphi\in Fm(C)$. (Substitution Property)
\end{itemize}
For all assignments $\gamma\colon V\rightarrow M$ and all formulas $\varphi,\psi\in Fm(C)$ the following \textbf{truth conditions} hold:
\begin{enumerate}
\item $\varGamma(\bot)\in M\smallsetminus \mathit{TRUE}$, $\varGamma(\top)\in \mathit{TRUE}$
\item $\gamma(\varphi\rightarrow\psi)\in \mathit{TRUE}\Leftrightarrow \gamma(\varphi)\notin \mathit{TRUE}\text{ or }\gamma(\psi)\in \mathit{TRUE}$
\item $\gamma(\neg\varphi)\in \mathit{TRUE}\Leftrightarrow \gamma(\varphi)\notin \mathit{TRUE}$
\item $\gamma(\varphi\wedge\psi)\in \mathit{TRUE}\Leftrightarrow \gamma(\varphi)\in \mathit{TRUE}$ and $\gamma(\psi)\in \mathit{TRUE}$
\item $\gamma(\varphi\vee\psi)\in \mathit{TRUE}\Leftrightarrow \gamma(\varphi)\in \mathit{TRUE}$ or $\gamma(\psi)\in \mathit{TRUE}$
\item $\gamma(\square\varphi)\in \mathit{TRUE}\Leftrightarrow \gamma(\varphi)\in \mathit{NEC}$
\item $\gamma(\varphi\equiv\psi)\in \mathit{TRUE}\Leftrightarrow \gamma(\varphi)=\gamma(\psi)$
\item $\gamma(\forall x\varphi)\in \mathit{TRUE}\Leftrightarrow\gamma_x^a(\varphi)\in \mathit{TRUE}$ for all $a\in M$
\item if $\gamma(\varphi\rightarrow\psi)\in \mathit{NEC}$, then $\gamma(\square\varphi\rightarrow\square\psi)\in \mathit{NEC}$
\item if $\gamma(\forall x\varphi)\in \mathit{NEC}$, then $\gamma_x^a(\varphi)\in \mathit{NEC}$ for all $a\in M$
\end{enumerate}
\end{definition}

The following Substitution Lemma II is a version of [Lemma 3.14, \cite{lewnd}].

\begin{lemma}[Substitution Lemma II]\label{990}
Let $\mathcal{M}$ be a simple model and $\varphi\in Fm(C)$. If $\sigma,\sigma'\colon V\rightarrow Fm(C)$ are substitutions and $\gamma,\gamma'\colon V\rightarrow M$ are assignments such that $\gamma(\sigma(x))=\gamma'(\sigma'(x))$ for all $x\in fvar(\varphi)$, then $\gamma(\varphi[\sigma])=\gamma'(\varphi[\sigma'])$.
\end{lemma}

The relation of satisfaction (truth) is defined as before, we use the same notation. Similarly as before, one verifies that a simple model satisfies all axioms of $\mathbb{AX}^-$ under any assignment (instead of the Substitution Lemma and the Coincidence Lemma now apply the Substitution Property and the Coincidence Property, respectively). In order to achieve soundness of the rule of Axiom Necessitation we impose the following semantic constraint:

\begin{definition}
Let $\mathcal{M}$ be a simple model with universe $M$ and the set of necessary propositions $\mathit{NEC}$. An assignment $\gamma\colon V\rightarrow M$ is called admissible if $\gamma(\varphi)\in \mathit{NEC}$ whenever $\varphi\in\mathbb{AX}^-$. $\mathcal{M}$ is called an admissible model if every assignment $\gamma\colon V\rightarrow M$ is admissible.
\end{definition}

Note that in an admissible (simple) model, $\mathit{NEC}\neq\varnothing$.

We write $\Phi\vdash\varphi$ if there is a derivation of $\varphi$ from $\Phi$ using axioms from $\mathbb{AX}^-$ and the rules of Modus Ponens and Axiom Necessitation. We write $\Phi\Vdash\varphi$ if for every admissible simple model $\mathcal{M}$ and any assignment $\gamma\colon V\rightarrow M$, $(\mathcal{M},\gamma)\vDash\Phi$ implies $(\mathcal{M},\gamma)\vDash\varphi$.

\begin{theorem}[Soundness and Completeness of $\mathbb{AX}^-$]\label{995}
For $\Phi\cup\{\varphi\}\subseteq Fm(C)$
\begin{equation*}
\Phi\vdash\varphi\Leftrightarrow\Phi\Vdash\varphi.
\end{equation*}
\end{theorem}

\paragraph*{Proof.}
We have already discussed soundness of the calculus and now concentrate on the completeness proof. The results and definitions \ref{485} -- \ref{595} of the first completeness proof remain unchanged. Of course, also the Deduction Theorem and Generalization can be adopted without any restrictions. Our task is now to construct an admissible simple model for a given set $\Phi$ which is a Henkin set w.r.t. the system based on $\mathbb{AX^-}$. The construction is very similar to that given in the proof of Theorem \ref{620}. The universe $M$, the sets $\mathit{TRUE}$ and $\mathit{NEC}$ and the Gamma-function are defined in the same way. We do not define operations $f_\top, f_\bot, f_\neg, f_\square, f_\rightarrow, f_\vee, f_\wedge, f_\equiv$ and $f_\forall$. Instead, we have to determine in which way an assignment $\gamma\colon V\rightarrow M$ extends to a function $\gamma\colon Fm(C)\rightarrow M$ such that the structural properties and the truth conditions of a simple model are satisfied. For a given assignment $\gamma\colon V\rightarrow M$ we fix a function $\tau_\gamma\colon V\rightarrow Fm(C)$ with the property $\tau_\gamma(x)\in\gamma(x)$ for every $x\in V$. The Claim 2 below shows that the actual choice $\tau_\gamma(x)\in \gamma(x)$ is not relevant. We interpret $\tau_\gamma$ as a substitution (this implies $\tau_\gamma(c)=c$ for $c\in C$). As in the first completeness proof, the relation $\approx_\Phi$ is defined by $\Phi\vdash\varphi\equiv\psi$, where $\Phi$ is maximally consistent, and by $\overline{\varphi}$ we denote the equivalence class of $\varphi$ modulo $\approx_\Phi$. Then we define the extension of an assignment $\gamma\colon V\rightarrow M$ by
\begin{equation*}
\gamma(\varphi):=\overline{\varphi[\tau_\gamma]},
\end{equation*}
for $\varphi\in Fm(C)$.\\
\textbf{Claim} 2: Let $\sigma,\sigma'\colon V\rightarrow Fm(C)$ be substitutions. If $\sigma(x)\approx_\Phi\sigma'(x)$ for all $x\in fvar(\varphi)$, then $\varphi[\sigma]\approx_\Phi\varphi[\sigma']$.\\
\textit{Proof of the Claim}: Let $fvar(\varphi)=\{x_1,...,x_n\}$. We may assume that $\sigma=[x_1:=\psi_1,...,x_n:=\psi_n]$ and $\sigma'=[x_1:=\psi'_1,...,x_n:=\psi'_2]$, and we may also assume that no $x_i$, $i=1,...,n$, occurs free in any of the $\psi_1,...,\psi_n,\psi'_1,...,\psi'_n$ (otherwise, we may replace such variables in $\varphi$ with others). Then the simultaneous substitutions $\sigma,\sigma'$ can be carried out successively. That is, applying successively axiom (vii) we obtain: $\varphi[\sigma]=\varphi[x_1:=\psi_1,...,x_n:=\psi_n]\approx_\Phi\varphi[x_1:=\psi_1,...,x_{n-1}:=\psi_{n-1}, x_n:=\psi'_n]\approx_\Phi\varphi[x_1:=\psi_1,...,x_{n-2}:=\psi_{n-2}, x_{n-1}:=\psi'_{n-1}, x_2:=\psi'_n]\approx_\Phi...\approx_\Phi\varphi[x_1:=\psi'_1,...,x_n:=\psi'_n]=\varphi[\sigma']$.\\
\textbf{Claim} 3: The structural conditions of a simple model are satisfied.\\
\textit{Proof of the Claim}: Clearly, $\gamma(c)=\overline{c}=\varGamma(c)$ for $c\in C$. In order to show the Coincidence Property let $\varphi\in Fm(C)$ and let $\gamma$, $\gamma'$ be assignments such that $\gamma(x)=\gamma'(x)$ for all $x\in fvar(\varphi)$. Then $\tau_\gamma(x)\approx_\Phi\tau_{\gamma'}(x)$ for all $x\in fvar(\varphi)$. Now we may apply Claim 2. Next, we show the Substitution Property. Let $\gamma\colon V\rightarrow M$ be an assignment, $\sigma\colon V\rightarrow Fm(C)$ a substitution and $\varphi\in Fm(C)$. We must show: $\gamma(\varphi[\sigma])=\gamma\sigma(\varphi)$. Recall that $\gamma\sigma\colon V\rightarrow M$ is the assignment given by $x\mapsto\gamma(\sigma(x))$, Definition \ref{310}. Then $\gamma(\varphi[\sigma])=\overline{\varphi[\sigma] [\tau_\gamma]}$ and $\gamma\sigma(\varphi)=\overline{\varphi[\tau_{\gamma\sigma}]}$. So it is enough to prove that $\varphi[\sigma] [\tau_\gamma] \approx_\Phi\varphi[\tau_{\gamma\sigma}]$. By induction on formulas one shows that for any $\chi\in Fm(C)$ and any substitutions $\sigma_1$ and $\sigma_2$: $\chi[\sigma_1][\sigma_2]=\chi[\sigma_1\circ\sigma_2]$, where $\sigma_1\circ\sigma_2$ is the substitution defined by $x\mapsto \sigma_1(x)[\sigma_2]$ (``first $\sigma_1$, then $\sigma_2$"). So it remains to show that $\varphi[\sigma\circ\tau_\gamma] \approx_\Phi\varphi[\tau_{\gamma\sigma}]$. Let $x\in fvar(\varphi)$. By definition, $(\sigma\circ\tau_\gamma)(x)=\sigma(x)[\tau_\gamma]$. On the other hand, $\tau_{\gamma\sigma}(x)\in\gamma\sigma(x)=\gamma(\sigma(x))=\overline{\sigma(x)[\tau_\gamma]}$. Hence, $(\sigma\circ\tau_\gamma)(x)\approx_\Phi\tau_{\gamma\sigma}(x)$, for all $x\in fvar(\varphi)$. The assertion now follows from Claim 2. Thus, the Substitution Property holds. \\
\textbf{Claim} 4: The truth conditions of a simple model are satisfied.\\
\textit{Proof of the Claim}: We show truth condition (iv). $\gamma(\varphi\wedge\psi)\in TRUE\Leftrightarrow(\varphi\wedge\psi)[\tau_\gamma]\in \Phi\Leftrightarrow\varphi[\tau_\gamma]\wedge\psi[\tau_\gamma]\in\Phi\Leftrightarrow\varphi[\tau_\gamma]\in\Phi$ and $\psi[\tau_\gamma]\in\Phi$. Most of the remaining truth conditions follow similarly applying axioms from $\mathbb{AX}^-$. We concentrate on the quantifier case:
\begin{equation*}
\begin{split}
&\gamma(\forall x\varphi)\in \mathit{TRUE}\\
&\Leftrightarrow (\forall x\varphi) [\tau_\gamma]\in\Phi\\
&\Leftrightarrow \forall y(\varphi[\tau_\gamma[x:=y]])\in\Phi,\text{ where } y\text{ is the forced variable}\\
&\Leftrightarrow\varphi[\tau_\gamma[x:=y]][y:=c]\in\Phi,\text{ for all }c\in C,\text{ since }\Phi\text{ is a Henkin set}\\
&\overset{(*)}{\Leftrightarrow}\varphi[\tau_\gamma[x:=c]]\in\Phi,\text{ for all }c\in C\\
&\overset{(**)}{\Leftrightarrow}\varphi[\tau_{\gamma_x^{\overline{c}}}]\in\Phi,\text{ for all }c\in C\\
&\Leftrightarrow\gamma_x^{\overline{c}}(\varphi)\in \mathit{TRUE},\text{ for all }\overline{c}\in M
\end{split}
\end{equation*}
It remains to show that the equivalences (*) and (**) hold.\\
(*): We have to ensure that $y\notin fvar(\varphi[\tau_\gamma])$. This follows from the fact that $y$ is the variable forced by substitution $\tau_\gamma$ w.r.t. $\forall x\varphi$.\\ 
(**): Let $z\in fvar(\varphi)$. First, we suppose $z\neq x$. Then $\tau_\gamma[x:=c](z)=\tau_\gamma(z)\in\gamma(z)$ and $\tau_{\gamma_x^{\overline{c}}}(z)\in\gamma_x^{\overline{c}}(z)=\gamma(z)$. Thus, $\tau_\gamma[x:=c](z)\approx_\Phi\tau_{\gamma_x^{\overline{c}}}(z)$. Now suppose $z=x$. Then $\tau_\gamma[x:=c](z)=c$ and $\tau_{\gamma_x^{\overline{c}}}(z)\in\gamma_x^{\overline{c}}(z)=\overline{c}$. Again, $\tau_\gamma[x:=c](z)\approx_\Phi\tau_{\gamma_x^{\overline{c}}}(z)$. By Claim 2, $\varphi[\tau_\gamma[x:=c]]\approx_\Phi\varphi[\tau_{\gamma_x^{\overline{c}}}]$. (**) now follows from Lemma \ref{595}.\\
Truth condition (x) follows similarly using the direction from left to right of the equivalence stated in Lemma \ref{530}.\\
\textbf{Claim} 5: $\mathcal{M}$ is an admissible (simple) model.\\
\textit{Proof of the Claim}: Let $\gamma\colon V\rightarrow M$ be an assignment. We show that $\gamma$ is admissible. Let $\varphi\in \mathbb{AX^-}$. By Axiom Necessitation, $\vdash_3\square\varphi$. Let $fvar(\varphi)=\{x_1,...,x_n\}$. By Lemma \ref{110}, $\vdash_3\forall x_1...\forall x_n\square\varphi$. Applying successively the axiom scheme (ix), we get $\psi:=\square\varphi[x_1:=c_1,...,x_2:=c_2]\in \Phi$, where the $c_i$ are constants with $c_i\approx_\Phi \gamma(x_i)$. By Claim 1 of Theorem \ref{620}, such constants exist. Moreover, $\gamma(c_i)=\overline{c_i}=\gamma(x_i)$. Now we apply Substitution Lemma II and the fact that $\psi\in \Phi$ contains no free variables and get: $\gamma(\square\varphi)=\gamma(\psi)=\overline{\psi[\tau_\gamma]}=\overline{\psi}\in \mathit{TRUE}$. By truth condition (vi), $\gamma(\varphi)\in \mathit{NEC}$.
Thus, 
\begin{equation*}
\mathcal{M}:=(M,\mathit{TRUE},\mathit{NEC},\varGamma)
\end{equation*}
is an admissible simple model. Consider now the canonical assignment $\beta\colon V\rightarrow M$ defined by $x\mapsto\overline{x}$.\\
\textbf{Claim} 6: $\varphi[\tau_\beta]\approx_\Phi\varphi$, for all $\varphi\in Fm(C)$.\\
\textit{Proof of the Claim}: We have $\tau_\beta(x)\approx_\Phi \varepsilon(x)$ for all $x\in fvar(\varphi)$, where $\varepsilon$ is the identity substitution. By Claim 2, $\varphi[\tau_\beta]\approx_\Phi\varphi[\varepsilon]$. By Lemma \ref{50}, $\varphi[\varepsilon]$ is alpha-congruent with $\varphi$. Alpha-congruence is contained in $\approx_\Phi$. Then the Claim follows from transitivity of $\approx_\Phi$.\\ 
Applying the definitions and Claim 6, we conclude: 
\begin{equation*}
(\mathcal{M},\beta)\vDash\varphi\Leftrightarrow\beta(\varphi)=\overline{\varphi[\tau_\beta]}=\overline{\varphi}\in \mathit{TRUE}\Leftrightarrow\varphi\in\Phi.
\end{equation*}
Hence, $(\mathcal{M},\beta)\vDash\Phi$. Finally, it remains to show that every consistent set extends to a Henkin set (in an extended language). We may adopt the construction given in the proof of Theorem \ref{640}. 
Q.E.D.

\begin{theorem}\label{1000}
Every $S3_\equiv^\forall$-model is an admissible simple model.
\end{theorem}

\paragraph*{Proof.}
Let $\mathcal{M}$ be a $S3_\equiv^\forall$-model. By the Coincidence Lemma \ref{300} and the Substitution Lemma \ref{320}, $\mathcal{M}$ has the Coincidence Property and the Substitution Property. Thus, the structural properties of a simple model are satisfied. The truth conditions follow from the truth conditions of a $S3_\equiv^\forall$-model along with the fact that every assignment $\gamma\colon V\rightarrow M$ of a $S3_\equiv^\forall$-model extends to a function on $Fm(C)$ such as specified in Definition \ref{200}. Since a $S3_\equiv^\forall$-model validates the rule of Axiom Necessitation, the model is also admissible. Q.E.D.\\

The converse of Theorem \ref{1000} is false. That is, the ``simple" semantics is strictly weaker or more general than the semantics of the first kind. This follows from the corresponding soundness and completeness theorems and the fact that $\mathbb{AX}^-$ is strictly contained in $\mathbb{AX}$. Nevertheless, given an admissible simple model $\mathcal{M}$, we are able to restore the structure of a Boolean prelattice on $M$. The function $f_\vee$, for instance, is defined as follows. Given any two elements $a,b\in M$, put $f_\vee(a,b):=\gamma(x\vee y)$ whenever $\gamma$ is an assignment and $x,y\in V$ such that $\gamma(x)=a$ and $\gamma(y)=b$. Of course, such an assignment and variables can be found. Moreover, that definition is independent of the particular assignment and the particular variables: Suppose there is another assignment $\gamma'$ and variables $u,v$ with $\gamma'(u)=a$ and $\gamma'(v)=b$. Let $\sigma=\varepsilon$ be the identity substitution and let $\sigma'$ be the substituition $[x:=u, y:=v]$. Then $\gamma(\sigma(x))=a=\gamma'(\sigma'(x))$ and $\gamma(\sigma(y))=b=\gamma'(\sigma'(y))$. Substitution Lemma II yields: $f_\vee(a,b)=\gamma(x\vee y)=\gamma((x\vee y)[\sigma])=\gamma'((x\vee y)[\sigma'])=\gamma'(u\vee v)$. However, it is not clear how to restore the higher-order function $f_\forall$ without a semantic property that corresponds to axiom (viii).\\

One goal of the paper was to present a non-Fregean semantics for some Lewis modal logics such that the relation of propositional identity does not suffer from too many restrictions. By the Collapse Theorem \ref{720}, propositional identity refines strict equivalence, and both relations collapse iff the given model is a Boolean algebra and satisfies the Collapse Axiom. The existence of an intensional model would imply that there are, up to alpha-congruence, no restrictions at all on the relation of propositional identity, more precisely, $\vdash_3\varphi\equiv\psi$ iff $\varphi=_\alpha\psi$, for all $\varphi,\psi\in Fm(C)$. The construction of an intensional model, however, is difficult because of the impredicativity of propositional quantification. We believe that a similar construction as in \cite{lewigpl} can be applied.

Finally, we would like to point out that our approach strongly relies on the modal principles inherent in Lewis modal systems $S3$--$S5$ and on the concept of propositional identity given by the axioms (v)--(vii). A non-Fregean semantics that captures $K$ as well as many other normal modal systems is found in \cite{ish1, ish2}. This is achieved by introducing a  concept of propositional identity which is axiomatized in a different way. However, the approach presented in \cite{ish1, ish2} involves the semantic limitations of standard modal logic: the Collapse Axiom is valid and models are Boolean algebras.

\end{document}